\documentclass[aps,amssymb,prb]{revtex4-1}
\usepackage{amsmath}
\usepackage{color}
\usepackage{graphicx}
\usepackage{hyperref}
\DeclareMathOperator\erf{erf}
\usepackage{dcolumn}
\usepackage{bm}
\usepackage{bigints}
\usepackage{natbib}
\usepackage{appendix}

\begin{document}

\newcommand{\revfirst}{}
\newcommand{\revsec}{}
\newcommand{\revnew}{}

\preprint{APS/123-QED}

\title{Plasmonic breathing modes in $\rm C_{60}$ molecules -- A quantum hydrodynamic approach} 

\author{Fatema Tanjia}
\affiliation{Universit\'e de Strasbourg, CNRS, Institut de Physique et Chimie des Mat\'eriaux de Strasbourg, UMR 7504, F-67000 Strasbourg, France}
\author{J\'er\^ome Hurst}
\affiliation{Universit\'e de Strasbourg, CNRS, Institut de Physique et Chimie des Mat\'eriaux de Strasbourg, UMR 7504, F-67000 Strasbourg, France}
\author{Paul-Antoine Hervieux}
\affiliation{Universit\'e de Strasbourg, CNRS, Institut de Physique et Chimie des Mat\'eriaux de Strasbourg, UMR 7504, F-67000 Strasbourg, France}
\author{Giovanni Manfredi}
\email[]{giovanni.manfredi@ipcms.unistra.fr}
\affiliation{Universit\'e de Strasbourg, CNRS, Institut de Physique et Chimie des Mat\'eriaux de Strasbourg, UMR 7504, F-67000 Strasbourg, France}

\date{\today}

\begin{abstract}
We propose and illustrate a quantum hydrodynamic (QHD) model for the description of plasmonic oscillations in the $\rm C_{60}$ molecule. Although simpler than competing approaches such as time-dependent density functional theory (TDDFT), the model contains the key ingredients to characterize plasmonic modes, namely the Hartree, exchange and correlation potentials, as well as nonlocal, nonlinear and quantum effects to the lowest order.
A variational technique is used to solve analytically the QHD model for the case of breathing (monopolar) plasmonic oscillations, revealing a bulk mode near the plasmon frequency. Numerical simulations of both the QHD equations and a TDDFT model confirm the existence of this mode and highlight a second collective mode at lower energy.
Such monopolar modes may be measured experimentally using electron energy loss spectroscopy.

\end{abstract}


\maketitle 

\section{Introduction}\label{sec:intro}
Recent years have witnessed a remarkable surge of interest in the electronic properties of nanomaterials, particularly when excited by ultrafast (femtosecond or shorter) pulses of electromagnetic radiation \cite{Pelton2008,Zhu2016}.
Nanoplasmonics may be defined as the study of the interactions of electromagnetic waves with the free electrons in a small object of nanometric dimensions, together with the collective phenomena that accompany such interactions \cite{Stockman2011,Manfredi2018}.
All sorts of nano-objects have been studied in the context of nanoplasmonics, including spherical and non-spherical nanoparticles, thin films, rods, and disks, as well as various assemblies of such objects (dimers, trimers, arrays, chains).
Hollow nano-objects deserve a special mention, as they will be the main topic of the present work. They include metallic nanoshells, as well as fullerene molecules \citep{Kroto1985} such as $\rm C_{60}$ and $\rm C_{240}$, or even nested fullerenes \cite{McCune2011}.

In metallic or metal-like nanomaterials, the valence electrons respond quickly to the external excitation, and begin to oscillate collectively at a well-defined resonant frequency. A typical example of such response is the localized surface plasmon (LSP) mode in a metallic nanoparticle \cite{Kreibig1995,Maier2007,Pelton2013}. The mode is  excited by an ultrafast laser pulse, usually in the visible range. The electric field of the laser drives the electrons away from their original steady state. The Coulomb force exerted by the ion lattice tends to bring the electrons back to equilibrium, but due to their inertia they overshoot it and begin to collectively oscillate at the so-called Mie frequency.

The LSP mode is by construction a dipole mode, because the electromagnetic wave length is much larger than the diameter of the nanoparticle.
Higher-order modes (quadrupole, octupole, \dots) were investigated in the recent past both theoretically \cite{Hoflich2009} and experimentally, by resorting to clever configurations in the optical experiments \cite{Krug2014}.
The plasmonic monopole (or breathing) mode is more difficult to excite due to its spherical symmetry. It is also harder to measure as it does not emit any electromagnetic radiation -- it is a {\em dark} mode \cite{Miao2017,Gomez2013}. Nevertheless, monopole modes have been observed  in silver
nanodisks using electron energy loss spectroscopy (EELS) \cite{Schmidt2012} or, more recently, optical spectroscopy \cite{Krug2014}.
{\revsec
EELS was also recently applied to collective plasmon excitations in $\rm C_{60}$ molecules \cite{Bolognesi2012,JPhysB141002}.}

{\revfirst $\rm C_{60}$ fullerene is a large molecule}
that displays some metallic properties due to the presence of delocalized electrons. Indeed, its 120 very tightly bound 1s electrons are often represented by means of a simplified jellium model, and only the dynamics of the remaining 240 delocalized valence electrons is treated self-consistently.
Compared to other nano-objects with similar geometry, such as the metallic nanoshells mentioned above, $\rm C_{60}$ is very small (diameter $\approx 0.7\rm\, nm$), hence it displays strong quantum and nonlocal features. For this reason, it may constitute an ideal arena to investigate typical quantum nanoplasmonic effects, which should be more prominent than in larger metallic nano-objects.
In particular, the $\rm C_{60}$ giant plasmonic oscillations observed at relatively high energy (20-40 eV) in the optical spectrum make it an attractive candidate for possible exciting applications, as well as for the fundamental understanding of the underlying physical effects.

Despite the fact that the $\rm C_{60}$ molecule is a very small nano-object,  its ab-initio description is a {\revsec complex} computational task.
Past theoretical and computational studies have used a variety of methods, ranging from Hartree-Fock (HF), \citep{Talapatra1992,Sheka2007} to the random phase approximation (RPA) \citep{Bertsch1991}, density functional theory (DFT) \citep{Scully2005, Madjet2008,Choi2017,PhysRevA.88.043201,Verkhovtsev2016}, and a Thomas-Fermi approach \citep{Palade2015}.
However, these approaches remain computationally costly, particularly when studying the dynamical properties beyond the linear response.

Quantum hydrodynamic (QHD) models \citep{Manfredi2001, Manfredi2005, Bonitz2018,Toscano2015,Ciraci2013,Ciraci2016} offer a useful and simpler alternative to ab-initio calculations for large N-body systems. In such models, the  electron dynamics is described by a small number of macroscopic fluid-like equations (continuity, Euler, energy conservation) that include quantum effects via the Bohm potential. The QHD approach can easily incorporate nonlocal and nonlinear effects, {\revsec structured jellium profiles}, as well as effects beyond the mean field approximation (exchange and correlations) along the same lines as time-dependent DFT (TDDFT). Indeed, QHD methods may be viewed as a particular class of time-dependent orbital-free DFT.
All in all,  QHD models, although sufficiently simple to run on a standard desktop computer, contain enough physics to study the full electron response well beyond the classical Mie theory.
Recent applications of QHD relevant to nanoplasmonics include surface plasmon modes in thin metal films \cite{Crouseilles2008}, metallic nanoparticles \citep{Hurst2014}, and semiconductor quantum wells \cite{Haas2009,Hurst2016}.

Here, we present some of the first applications of QHD to plasmonic breathing modes in $\rm C_{60}$ molecules. After illustrating the basic QHD equations in Sec. \ref{sec:model}, we develop a variational method that allows us to reduce the macroscopic electron dynamics to a single effective ordinary differential equation (Sec. \ref{sec:lagrangian}), which is then used to evaluate the linear and nonlinear dynamics of the plasmonic breathing modes. Finally, in Sec. \ref{sec:numerical} the results are compared to  numerical simulations of the full QHD equations and to linear-response theory using a TDDFT approach.

\section{QHD modelling of $\rm C_{60}$} \label{sec:model}

In this Section, we provide a short derivation of the QHD equations for the particular case of $\rm C_{60}$; more details can be found in our earlier works \cite{Manfredi2001, Manfredi2005}.

As mentioned in the introduction, we adopt a jellium model that takes into account the 120 1s localized electrons. The remaining 240 valence electrons can be represented by the time-dependent Kohn-Sham equations (atomic units are used throughout this work):
\begin{equation}
i \frac{\partial\psi_l}{\partial t} =\left( -\frac{1}{2}\Delta_r-V_H + \frac{l(l+1)}{2r^2} + V_{X,C}+V_{ps}\right) \psi_l \, \label{eq:ks}
\end{equation}
where we assumed radial symmetry from the start, so that $\Delta_r = r^{-2}\partial_r r^2 \partial_r$ stands for the radial part of the Laplacian. The various terms represent respectively the Hartree potential $V_H$, the centrifugal potential $V_l$ ($l$ is the azimuthal quantum number),  exchange and correlations $V_{X,C}$, and a pseudopotential $V_{ps}$ commonly employed in the DFT literature in order to recover the correct ionization potential for $\rm C_{60}$.
The Hartree potential is a solution of the Poisson equation
\begin{equation}
\Delta_r V_H=4\pi \left(\sum_{l} p_l |\psi_l|^2-n_i\right), \label{eq:poisson}\
\end{equation}
where the $p_l$ are the occupation numbers and $n_i(r)$ is the ion jellium density.

We have chosen the wave function normalization in such a way that
\begin{equation}
N=\sum_{l} p_l\int_0^\infty 4\pi r^2|\psi_l|^2dr=\int_0^\infty 4\pi r^2 n_i dr,
\label{eq:normalization}
\end{equation}
where $N=\sum_{l} p_l =240$ is the total number of valence electrons.
Note that the sum extends over both $\sigma$ and $\pi$ electrons, which are characterized by different radial quantum numbers ($n_r=0$ for the former and $n_r=1$ for the latter). The $\rm C_{60}$ ground-state configuration, obtained from a full DFT calculation, is summarized in the Appendix \ref{A1}.

To obtain the QHD equations, first we make a Madelung transformation on the radial wave function
\begin{equation}\label{eq:madelung}
\psi_l(r,t)=A_l \exp(i S_l),
\end{equation}
where the real amplitudes $A_l$ and phase $S_l$ are related to the density $n_l$ and velocity $u_l$ of each wave function through:
\begin{eqnarray}
&&n_l(r,t)=|\psi_l|^2=A_l^2, \label{4a}\\
&&u_l(r,t)= \partial_r S_l . \label{4b}
\end{eqnarray}
Substituting Eq. \eqref{eq:madelung} into Eq. \eqref{eq:ks} and separating the real and imaginary parts, we get
{\small\begin{eqnarray}
&& \frac{\partial n_l}{\partial t} + \frac{\partial}{\partial r}(n_l u_l)+\frac{2}{r}\,n_l u_l=0 \,, \label{5a}\\
&& \frac{\partial u_l}{\partial t}+u_l\frac{\partial u_l}{\partial r}   = \frac{\partial V_H}{\partial r}+\frac{1}{2}\frac{\partial}{\partial r} \left(\frac{\Delta_r\sqrt{n_l}}{\sqrt{n_l}}\right)- \frac{\partial}{\partial r}\left[\frac{l(l+1)}{2r^2}\right] -\frac{\partial V_{X,C}}{\partial r}-\frac{\partial V_{ps}}{\partial r}. \label{5b}\
\end{eqnarray}}
Now multiplying Eqs.(\ref{5a})-(\ref{5b}) by $p_l$ and summing over $l$, we get the following set of fluid equations:
\begin{eqnarray}
&& \frac{\partial n}{\partial t} + \frac{\partial}{\partial r}(n u)+\frac{2}{r}\,n u=0 \,, \label{8a1}\\
&& \frac{\partial u}{\partial t}+u\frac{\partial u}{\partial r}  = \frac{\partial V_H}{\partial r}+\frac{1}{2}\frac{\partial}{\partial r} \left(\frac{\Delta_r\sqrt{n}}{\sqrt{n}}\right)-\frac{1}{n}\frac{\partial P}{\partial r}-\frac{2}{r}\frac{P}{n}
-\frac{\partial V_L}{\partial r}-\frac{\partial V_{X,C}}{\partial r}-\frac{\partial V_{ps}}{\partial r}, \label{8a2}\
\end{eqnarray}
where
\begin{eqnarray}
&& n(r,t)=\sum_{l}p_l n_l  \,, \label{6a}\\
&& u(r,t)\equiv \langle u_l\rangle=\sum_{l} p_l \,\frac{n_l}{n}\, u_l, \label{6b}\\
&&P= n\left[\frac{\sum_l p_l\, n_l \,(u_l)^2}{n}-\left(\frac{\sum_{l} p_l \,n_l\, u_l}{n}\right)^2\right]= n\left[\langle u_l^2\rangle-\langle u_l\rangle^2\right],  \label{6c}\\
&& V_L=\frac{\langle L^2\rangle}{2r^2} = {1 \over N}\frac{\sum_l p_l\,l(l+1)}{2r^2} \label{6d}
\end{eqnarray}
are, respectively, the fluid electron density, the mean electron velocity, the electron pressure, and the average centrifugal potential. Using the values of the occupation numbers given in the Appendix \ref{A1}, one obtains $\langle L^2\rangle \approx 37.5$ in atomic units.
In terms of the fluid variables, Poisson's equation for the Hartree potential reads as:
\begin{equation}
\Delta_r V_H = 4\pi \left(n-n_i\right) \label{eq:poissonfluid}.
\end{equation}

To this point, the derivation of the QHD equations is exact, except for the second term on the right hand side of Eq. \eqref{8a2} (the so-called Bohm potential), which describes quantum effect to lowest order. To obtain this term, we had to assume that
\[
\left\langle\frac{\Delta_r\sqrt{n_l}}{\sqrt{n_l}}\right\rangle \approx  \frac{\Delta_r\sqrt{n}}{\sqrt{n}} ,
\]
which is approximately correct as long as spatial gradients are not too large \cite{Manfredi2001, Manfredi2005}. It can be recognized that the Bohm potential corresponds to the von Weizs{\"a}cker correction to the electron kinetic energy.

We still have to specify the electron pressure and the exchange-correlation potential.
For the time being, we neglect correlations, i.e. $V_{C}=0$, and use the local density approximation (LDA) for the exchange potential:
\begin{equation}\label{g1}
V_X [n]= -\frac{(3\pi^2)^{1/3}}{\pi}\, n^{1/3}.
\end{equation}
We further consider that the system's temperature is always much lower than the Fermi temperature of $\rm C_{60}$, so that the pressure can be approximated by that of a fully degenerate electron  gas:
\begin{equation}\label{g11}
P= {1 \over 5}(3\pi^2)^{2/3} n^{5/3}.
\end{equation}

{\revsec
As in most earlier studies\cite{Puska1993,Prodan2002}, the pseudopotential is taken to be constant inside the ionic jellium, with $V_{ps}=-0.7$, and zero elsewhere.
A more sophisticated structured pseudopotential was suggested recently \cite{JPhysB215101}, which is computed as the difference between the total potential obtained from an {\it ab initio} calculation and the one obtained from a pure jellium model. Such improved pseudopotential produces a more accurate ground-state electron density and energy levels, but requires a prior {\it ab initio} calculation to be implemented.
This and other structured pseudopotentials could be easily incorporated in the QHD model described here, and would presumably also improve the accuracy of the QHD calculations.
}

A final consideration is in order here concerning the pressure and angular momentum terms in Eqs. \eqref{8a1}-\eqref{8a2}, because some cancellations take place. If we consider a classical spherically-symmetric system in a generic potential $V(r)$, the radial component of the Euler equation of motion reads as
\begin{equation}\label{eq:classical}
\frac{\partial u}{\partial t}+u\frac{\partial u}{\partial r} = -\frac{\partial V}{\partial r} -{1\over n} \frac{\partial P_{rr}}{\partial r} - \frac{2}{r}\, \frac{P_{rr}-P_{t}}{n},
\end{equation}
where $P_{rr}$ is the radial part of the pressure and $P_t \equiv P_{\vartheta\vartheta}=P_{\phi\phi}$ is the tangential component, which  in a spherically symmetric system is identical for both angular coordinates $\vartheta$ and $\phi$.
Now, the tangential component of the pressure is related to the average angular momentum through:
\begin{equation}\label{eq:angmom}
\langle L^2 \rangle =  m^2 r^2 \langle v_t^2 \rangle = 2r^2 P_t/n,
\end{equation}
where we used the relation $\langle v_t^2 \rangle =2 P_t/(mn)$.
Using the above expression, one can readily show that $-\partial_r V_L = 2P_t/(n r)$: in other words, the term containing the tangential part of the pressure in the classical equation \eqref{eq:classical} is the same as the angular momentum term given by Eq. \eqref{6d}. Now, if we assume that the pressure is completely isotropic, i.e. $P_{rr}=P_t \equiv P$, then the angular momentum term exactly cancels the term $2P/(n r)$ in Eq. \eqref{8a2}.
{\revfirst This isotropy assumption is consistent with the study of spherically-symmetric monopole modes as envisaged here (although not necessarily with higher-order dipole and multipole modes) and will be adopted throughout the present work.}

\section{Variational approach for the QHD equations}\label{sec:lagrangian}

\begin{table}
  \centering
\begin{tabular}{|c|c|}
\hline
$\sigma_0$ &  $1.72\,a_0$\\
\hline
$\Omega$ & $1.24$ au $\approx 33.81$ eV\\
\hline
$\omega_p=\sqrt{4\pi n_0}$ &  $1.36$ au $\approx 37.1$ eV \\
\hline
$\omega_p/\sqrt{3}=\sqrt{4\pi n_0/3}$ & $0.787$ au $\approx 21.4$ eV \\
\hline
$r_{s}$ & $ 1.173\,a_0$\\
\hline
$R$ & $ 6.69\,a_0$\\
\hline
$\Delta$ & $ 2.84\,a_0$\\
\hline
$R_1$ & $ 5.27\,a_0$\\
\hline
$R_2$ & $ 8.11\,a_0$\\
\hline
$V$ & $ 1621\,a_0^3$\\
\hline
$n_{eq}=N/V$ & $ 0.15\,a_0^{-3}$\\
\hline
\end{tabular}
\caption{Ground state width $\sigma_0$ and linear response frequency $\Omega$ obtained from the QHD variational approach. Other relevant parameters are also summarized here.}
\label{tab:groundstate}
\end{table}

It can be shown  \cite{Hurst2014} that the set of fluid equations \eqref{8a1}-\eqref{8a2} together with Poisson's equation \eqref{eq:poissonfluid} can be exactly represented by a Lagrangian density $\mathcal{L}(n,\theta,V_H)$, where the function $\theta$ is related to the mean electron velocity, $u=\partial\theta/\partial r$. The expression for this Lagrangian density is the following:
{\small\begin{eqnarray}
&&\mathcal{L}=n\left[\frac{1}{2}\left(\frac{\partial\theta}{\partial r}\right)^2+\frac{\partial\theta}{\partial t}\right]+\frac{1}{8n}\left(\frac{\partial n}{\partial r}\right)^2+\frac{3}{10}(3\pi^2)^{2/3}n^{5/3}\nonumber\\
&& -\frac{3}{4\pi}(3\pi^2)^{1/3}n^{4/3}+nV_{ps}-(n-n_i)V_H-\frac{1}{8\pi}\left(\frac{\partial V_H}{\partial r}\right)^2. \label{eq:lagrangian}\
\end{eqnarray}}
By taking the Euler-Lagrange equations with respect to the fields $n$, $\theta$ and $V_H$, one recovers exactly the the fluid equations (\ref{8a1})-(\ref{8a2}) as well as  Poisson's equation \eqref{eq:poissonfluid} .

The idea here is to guess a ``reasonable" profile for the electron density, insert it into Eq. \eqref{eq:lagrangian}, and integrate over all space. Following recent numerical calculations of the $\rm C_{60}$ ground state \cite{Madjet2008,Palade2015}, a good approximation of the electron density is the following:
\begin{equation}
n(r,t)=\frac{A}{{\sigma(t)}^3}\left[\frac{r}{\sigma(t)}\right]^{k} \exp\left[-\frac{r^2}{2\sigma(t)^2}\right],\label{eq:ansatz}\
\end{equation}
where $\sigma(t)$ is a time-dependent variable representing the size of the electron cloud, $k$ is any positive even integer, and $A$ is a normalization constant. Using the normalization condition of Eq. \eqref{eq:normalization} we obtain
$$A=\frac{N\,2^{-k/2}}{4\sqrt{2}\,\pi\Gamma \left( \frac{k+3}{2} \right)} \,,$$
where $\Gamma$ is the Euler gamma function.
We found that a good match between the Ansatz of Eq. \eqref{eq:ansatz} and DFT calculations was obtained with $k=14$.

The ion density is given by the following expression
\begin{equation}
n_i(r)=n_{eq} \left[\mathcal{H}\left(r-R_1\right)-\mathcal{H}\left(r-R_2\right)\right]\, .\label{13c}\
\end{equation}
Here, $\mathcal{H}\left(r-R_1\right)$ and $\mathcal{H}\left(r-R_2\right)$ are Heaviside step functions, $R_1=R-\Delta/2$ and $R_2=R+\Delta/2$ are the inner and outer radii of the ionic jellium, where $R= 6.69$ is the average radius and $\Delta= 2.84$ is the width of the ion density, $n_{eq}=N/V\approx 0.15$ is the homogeneous positive charge density, and $V=4\pi\left(R_2^3-R_1^3\right)/3$ is the volume of the spherical shell occupied by the ions \cite{Madjet2008}.
A summary of the above parameters can be found in Table \ref{tab:groundstate}.

We now need to express the other variables ($\theta$ and $V_H$) in terms of the electron density $n$. The mean velocity $u$ can be obtained exactly from the continuity equation as $u=\left(\dot{\sigma}/\sigma\right)\,r$, where the dot denotes differentiation with respect to time.
From the relationship between $\theta$ and the mean velocity $u=\partial\theta/\partial r$,  we obtain $\theta=\left(\dot{\sigma}/2\sigma\right)\,r^2$.

For the Hartree potential, one needs to solve the Poisson equation with a source given by the electron density defined in Eq. \eqref{eq:ansatz}. This is done by separating the Hartree potential in two terms that pertain respectively to the ions and the electrons $V_H=V_i+V_e$. Extensive details of these calculations are provided in the Appendix \ref{A2}.

Finally, the Lagrangian function is obtained by integrating $\mathcal{L}$ over the whole space:
\begin{equation}
L(\sigma,\dot{\sigma})=\frac{1}{N}\int \mathcal{L}\,d\mathbf{r}=\frac{4\pi}{N}\int_0^{\infty} \mathcal{L}\,r^2dr, \label{a1}\
\end{equation}
which yields, after much algebra (see Appendix \ref{A2}):
\begin{equation}\label{18b}\
L(\sigma,\dot{\sigma})=-\frac{17\dot{\sigma}^2}{2}+\widetilde U(\sigma),
\end{equation}
where the potential $\widetilde U(\sigma)$ is given by the following function:
\begin{eqnarray}\label{18c}\
&&\widetilde U(\sigma)=\frac{\alpha_1}{\sigma^2}+\frac{\alpha_2N^{2/3}}{\sigma^2}-\frac{\alpha_3N^{1/3}}{\sigma}+\frac{\alpha_4 N}{\sigma}\nonumber\\
&&+\frac{R_1}{2027025}\,\exp\left(-\frac{R_1^2}{2\sigma^2}\right)\Biggl[-\sqrt{\frac{2}{\pi}}\,V_0\,\frac{F_1(R_1,\sigma)}{\sigma^{15}}\nonumber\\
&&+2\sqrt{2\pi}\,n_0\,\frac{K_1(R_1,R_2,\sigma)}{\sigma^{11}}\Biggr]-\frac{R_2}{2027025}\,\exp\left(-\frac{R_2^2}{2\sigma^2}\right)\nonumber\\
&&\Biggl[-\sqrt{\frac{2}{\pi}}\,V_0\,\frac{F_2(R_2,\sigma)}{\sigma^{15}}+2\sqrt{2\pi}\,n_0\,\frac{K_2(R_1,R_2,\sigma)}{\sigma^{11}}\Biggr]\nonumber\\
&&+\erf\left(\frac{R_1}{\sqrt{2\sigma}}\right)\Biggl[V_0-\frac{2\pi n_0}{3}K_3(R_1,R_2,\sigma)\Biggr]\nonumber\\
&&-\erf\left(\frac{R_2}{\sqrt{2\sigma}}\right)\Biggl[V_0-\frac{2\pi n_0}{3}K_4(R_1,R_2,\sigma)\Biggr],
\end{eqnarray}
for which $F_1(R_1,\sigma)$, $F_2(R_2,\sigma)$, $K_1(R_1,R_2,\sigma)$, $K_2(R_1,R_2,\sigma)$, $K_3(R_1,R_2,\sigma)$ and $K_4(R_1,R_2,\sigma)$ are given in Appendix \ref{A2}, $\alpha_1\approx 0.258$, $\alpha_2\approx 0.045$, $\alpha_3\approx 0.091$, $\alpha_4\approx 0.114$, and $\erf$ denotes the error function.

\begin{figure}
  \centering
  \includegraphics[width=0.4\paperwidth]{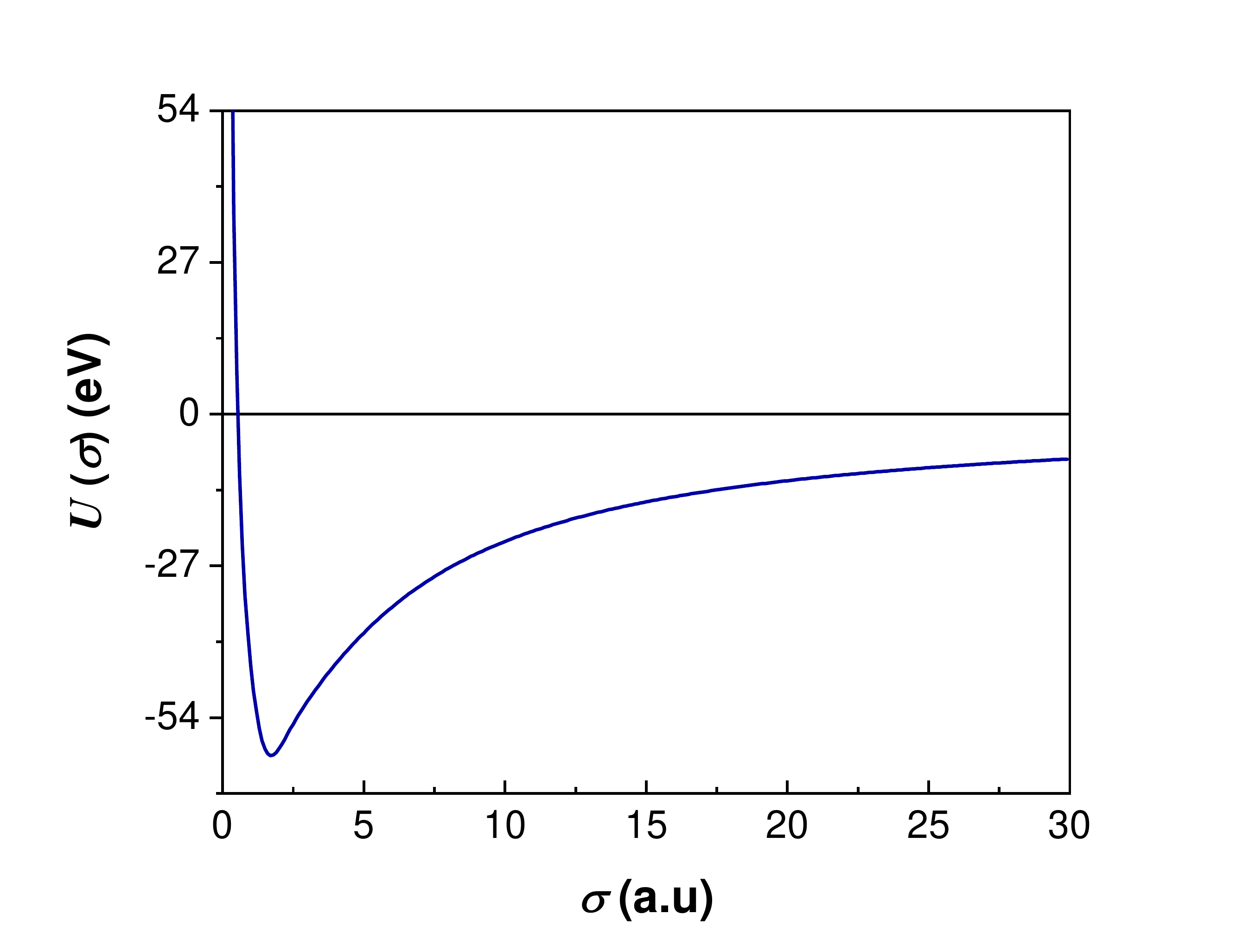}\\
  \caption{Profile of the effective potential $U(\sigma)$ as a function of $\sigma$}\label{fig:upoten}
\end{figure}

\begin{figure}
  \centering
  \includegraphics[width=0.4\paperwidth]{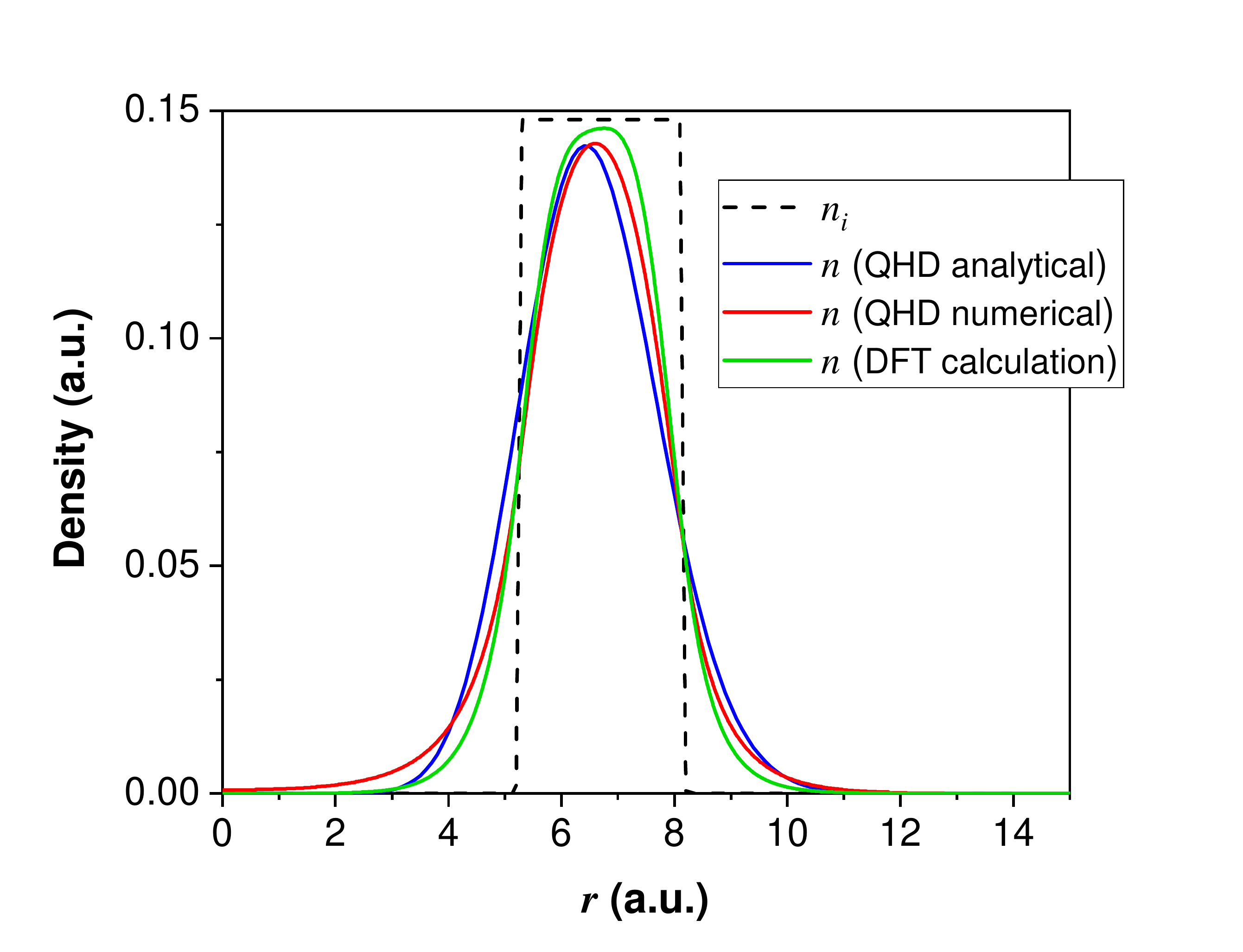}\\
  \caption{Ground-state electron density profiles (solid curves) obtained with different methods: QHD analytical (variational approach; blue line), QHD numerical (red) and DFT (green). The dashed curve represents the ion density.}\label{fig:densprofile}
\end{figure}

\begin{figure}
  \centering
  \includegraphics[width=0.4\paperwidth]{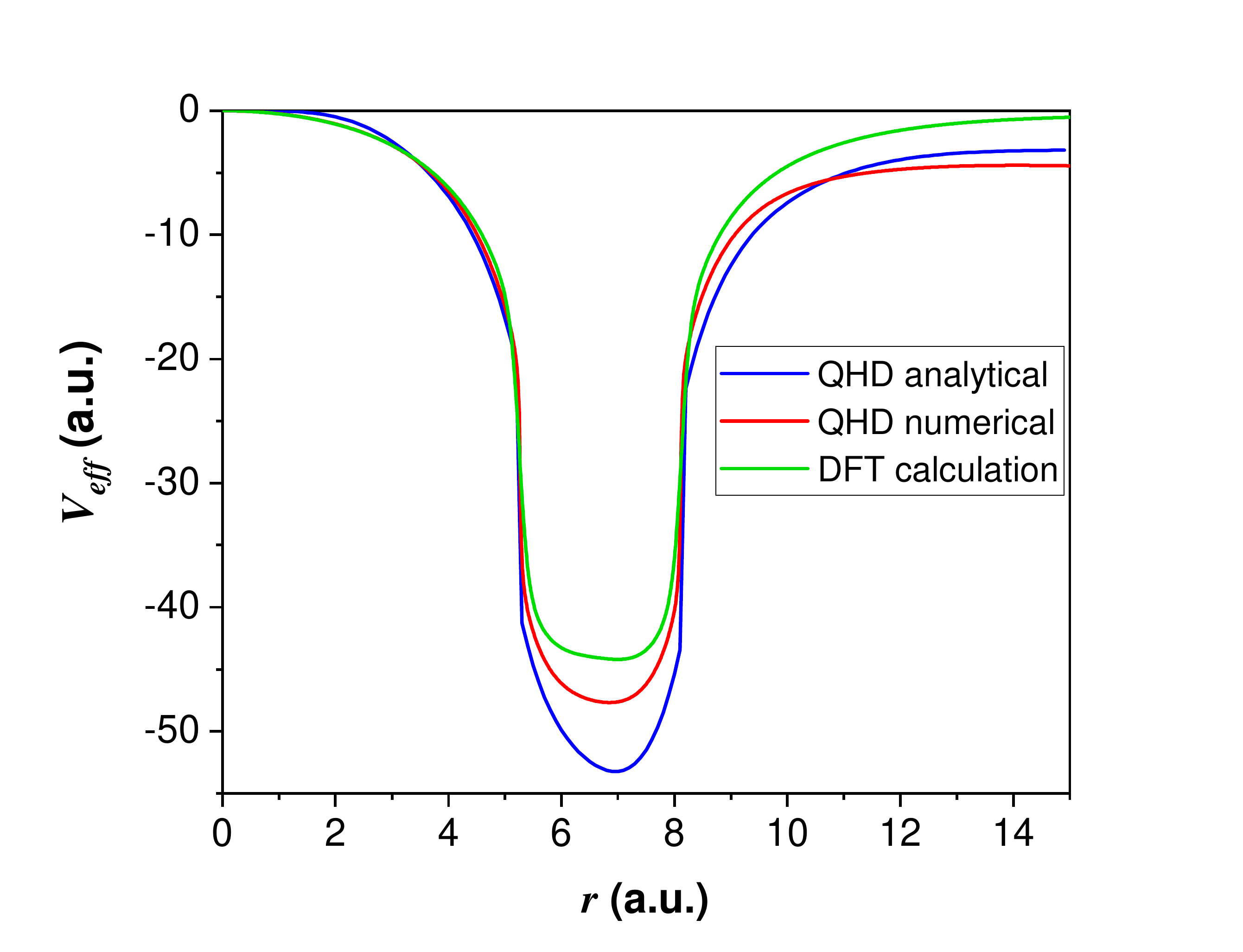}\\
  \caption{Effective (total) potentials corresponding to the three electron densities shown in Fig. \ref{fig:densprofile}. }\label{fig:veff}
\end{figure}

\begin{figure}
  \centering
  \includegraphics[width=0.4\paperwidth]{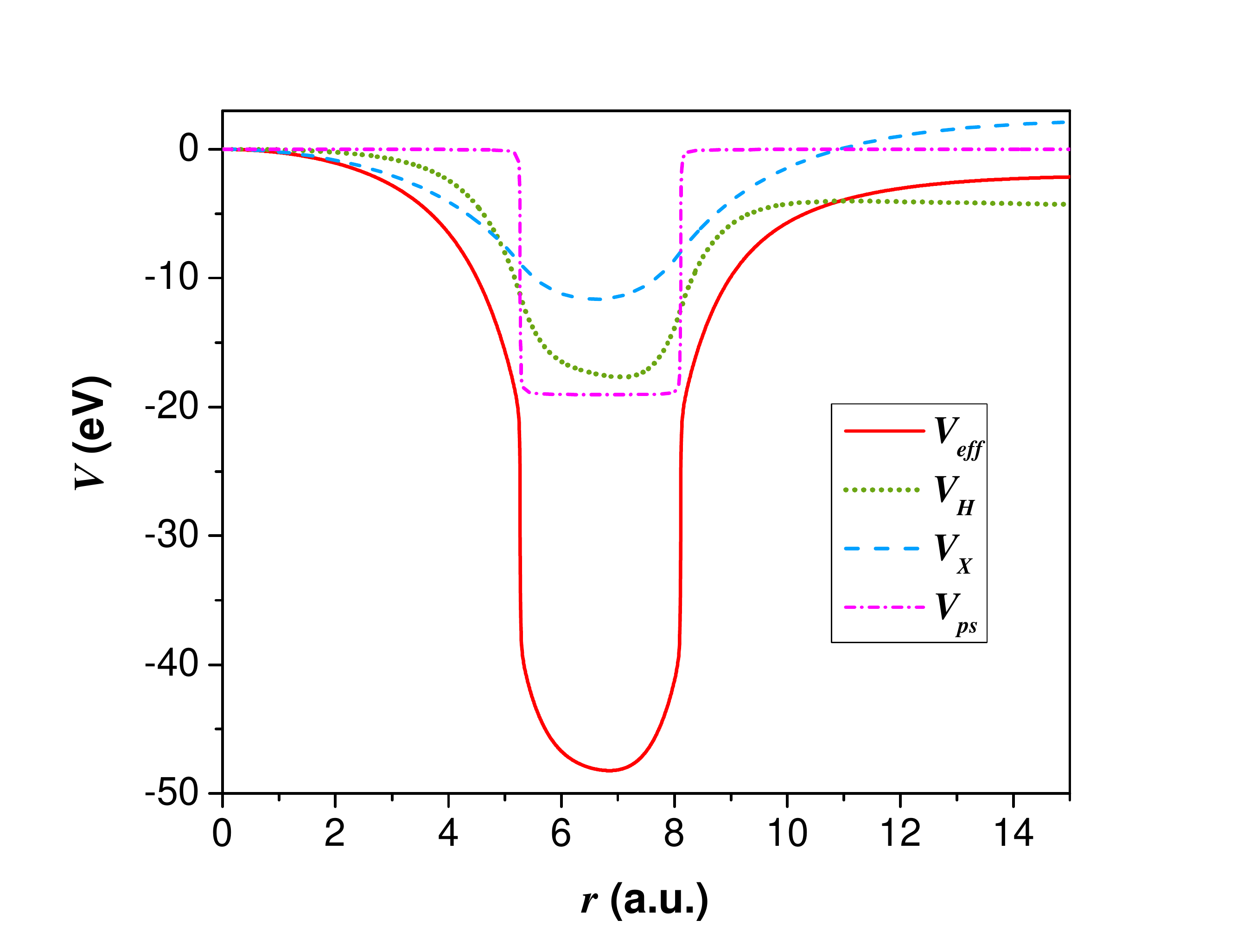}\\
  \caption{Various potential terms (Hartree, exchange, pseudopotential) obtained from the numerical solution of the QHD equations for the ground state. $V_{eff}$ is the sum of all these terms.}\label{fig:allpoten}
\end{figure}


The equation of motion of the system can be obtained from the Euler-Lagrange equation $\partial L/\partial\sigma-d/dt\left(\partial L/\partial\dot{\sigma}\right)=0$, which yields
\begin{equation}
\ddot{\sigma}=-\frac{\partial U}{\partial\sigma},
\label{eq:motion}
\end{equation}
where $U(\sigma)=\widetilde U(\sigma)/17$.
Despite the complicated form of $U(\sigma)$, the equation of motion for the width $\sigma$ of the electron cloud is rather simple, and resembles that of a fictitious particle evolving in an external potential. In the remaining part of this section, we will use Eq. \eqref{eq:motion} to deduce some ground-state and linear-response properties of the system.

\subsection{Ground state}
The profile of the potential $U(\sigma)$ is shown in Fig. \ref{fig:upoten}. It displays a single minimum located at $\sigma_0\approx 1.719$. Injecting this value into Eq. \eqref{eq:ansatz} we obtain the ground-state electron density of the system, which is plotted in Fig. \ref{fig:densprofile}. For comparison, we also plot the density profile obtained from the numerical solution of the full QHD equations, as well as the  density computed using a standard DFT code \cite{Maurat2009}. All parameters are the same for the three curves shown in Fig. \ref{fig:densprofile}.
This result shows that the parametrization given by Eq. \eqref{eq:ansatz} is a rather satisfactory one and may be used in different contexts to represent in a simple way the ground-state electron density of $\rm C_{60}$.
The corresponding effective potentials $V_{eff}=V_H+V_X+V_{ps}$ for these three different approaches are plotted in Fig. \ref{fig:veff}, also showing good agreement between them.
The various components of the effective potential, obtained from a numerical solution of the full QHD equations, are shown in Fig. 4.

\subsection{Linear and nonlinear response}
The linear response of the system can be computed analytically from our Lagrangian model. The frequency of linear oscillations around the minimum of the potential $U(\sigma)$ is given by
\[
\Omega=\sqrt{\,\left|U^{\prime\prime}(\sigma_0)\right|},
\]
where the apex denotes differentiation with respect to $\sigma$. One obtains that $\Omega \approx 33.8\,\rm eV$, which should be compared to the plasmon frequency $\omega_p \approx 37.1\,\rm eV$. For such a small system as $\rm C_{60}$, it is not surprising that the computed frequency is redshifted with respect to $\omega_p$, just like the localized surface plasmon frequency in a metallic nanoparticle \cite{Brechignac1993}.
The principal source of this redshift lies in the large spillout of the electron density with respect to the ion density, as is apparent in Fig. \ref{fig:densprofile}.
The parameters of the ground state and linear response are summarized in Table \ref{tab:groundstate}.

\begin{figure}
  \centering
  \includegraphics[width=0.4\paperwidth]{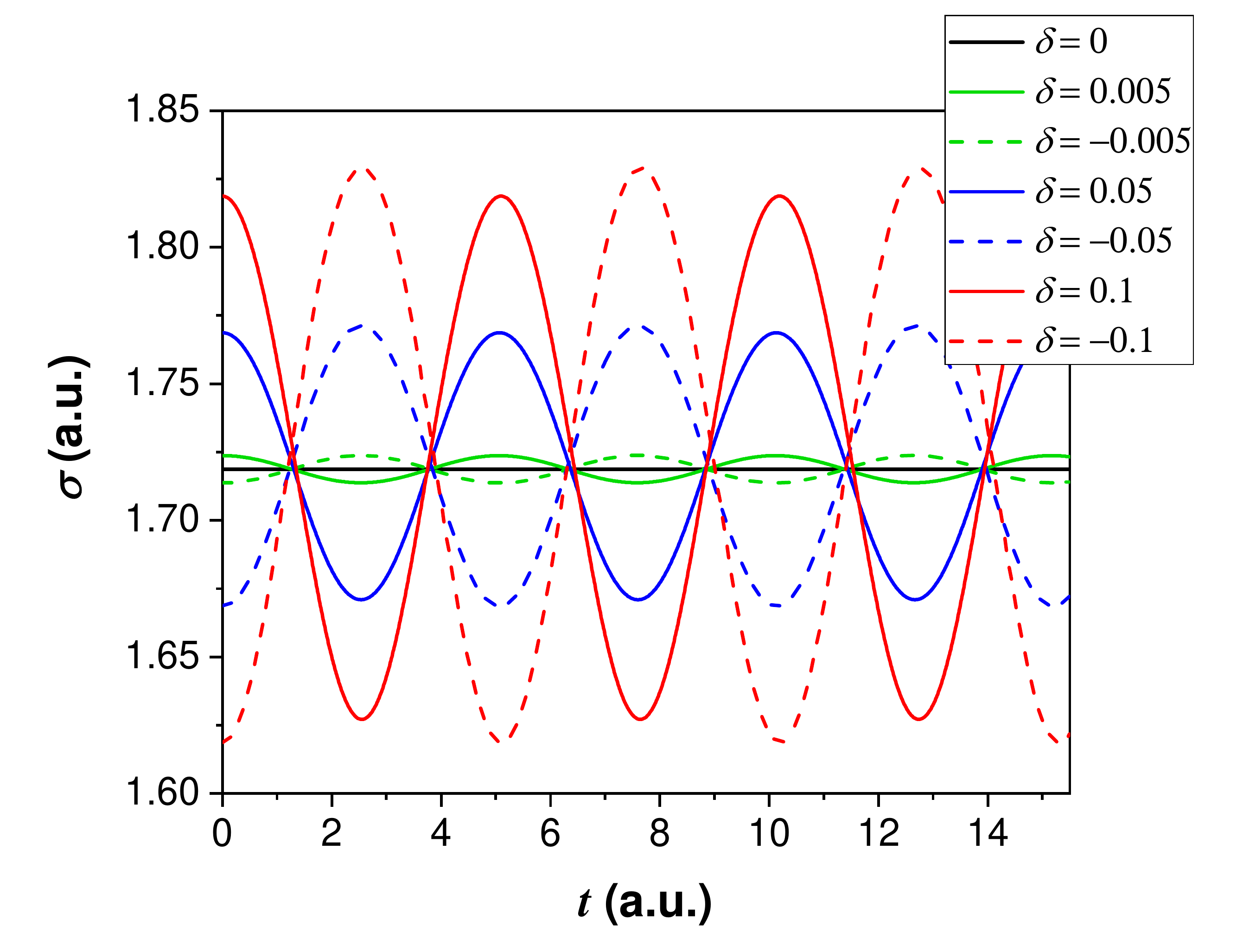}\\
  \caption{Breathing oscillations of the width $\sigma(t)$ around its equilibrium value $\sigma_0$ for different values of the excitation $\delta$.}\label{s1}
\end{figure}

\begin{figure}
  \centering
  \includegraphics[width=0.4\paperwidth]{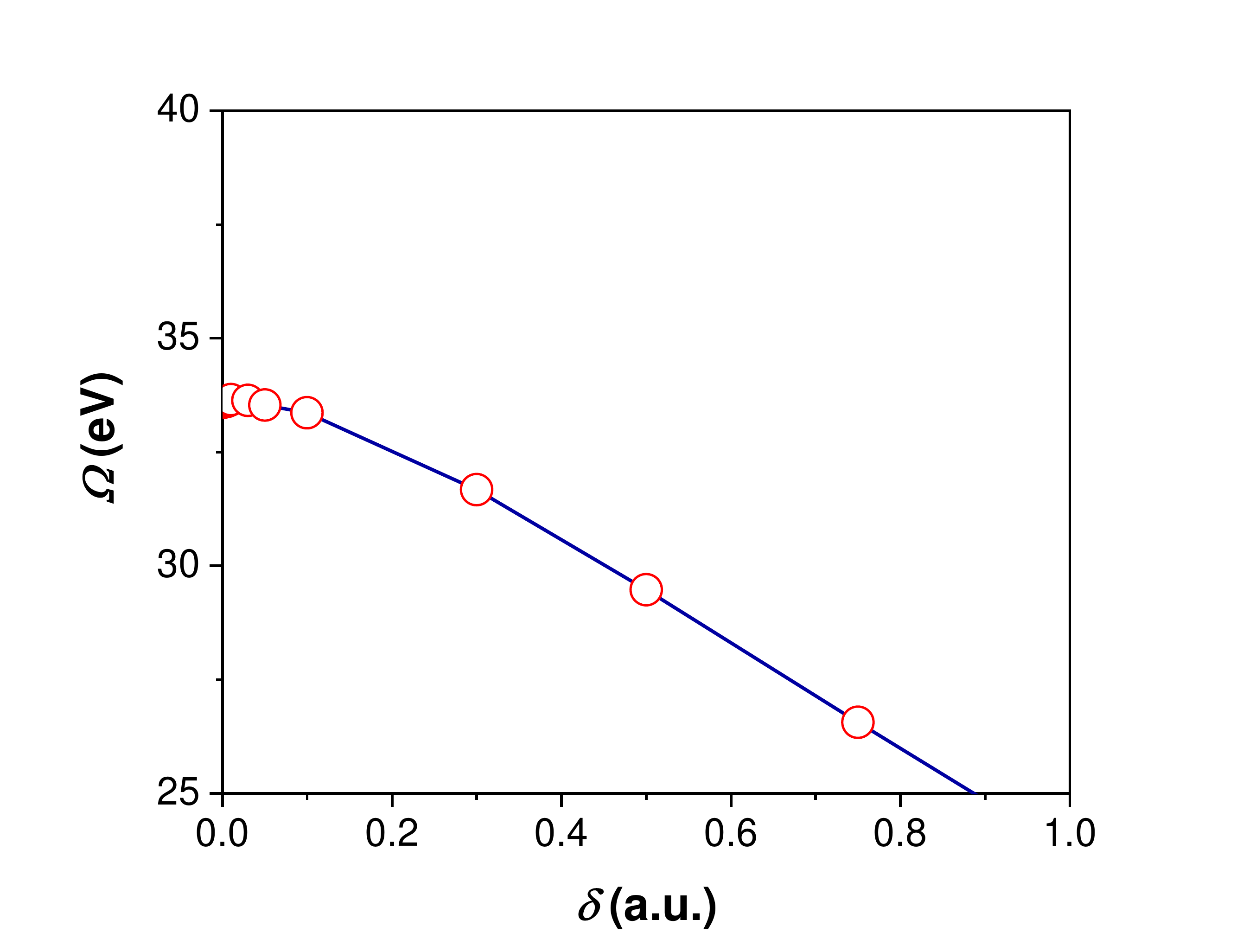}\\
  \caption{Observed oscillation frequency $\Omega$ for different values of the excitation $\delta$.}\label{f2}
\end{figure}

\begin{figure}
  \centering
  \includegraphics[width=0.4\paperwidth]{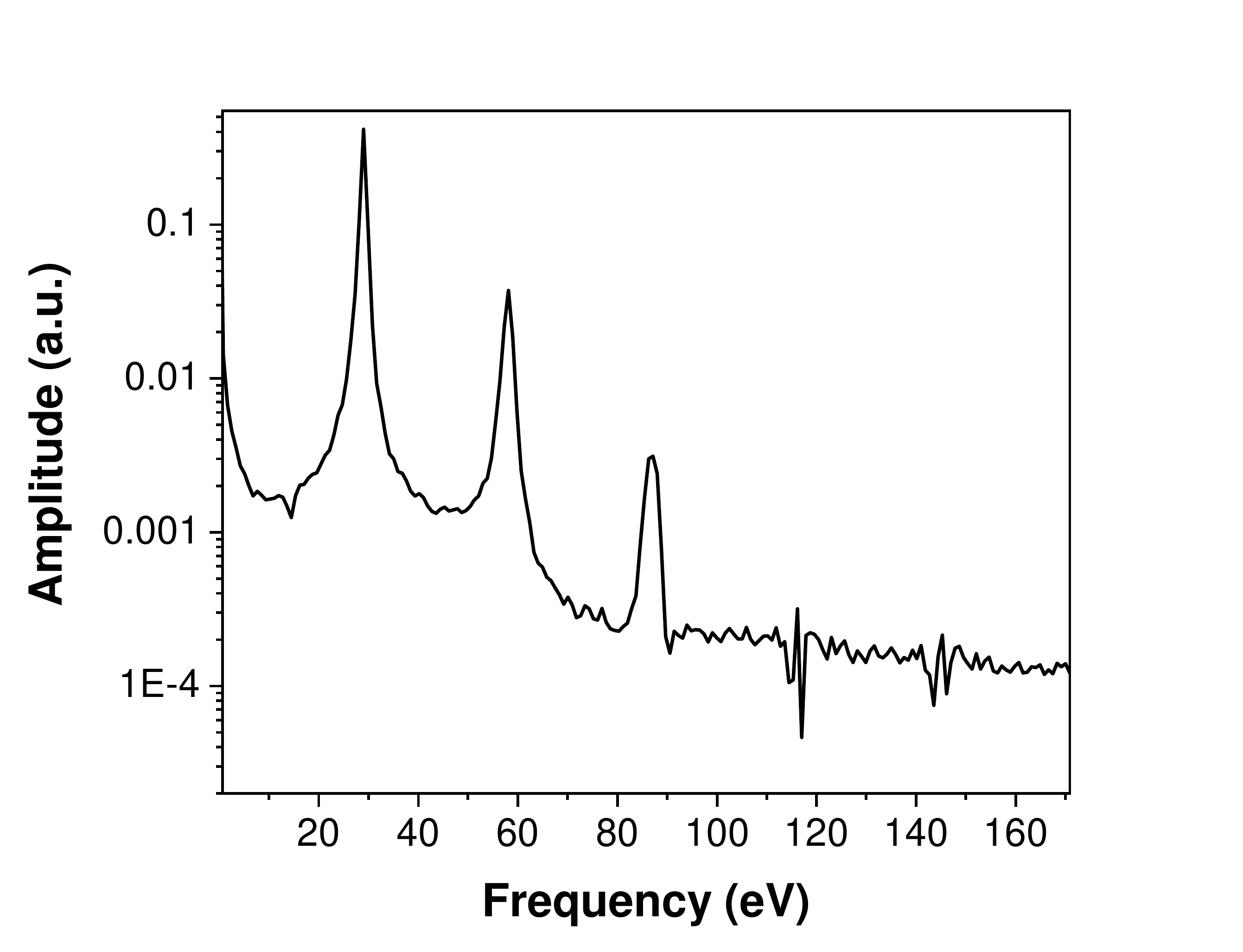}\\
  \caption{Frequency spectrum of the signal $\sigma(t)$ in a strongly nonlinear regime, $\delta=0.5$.}\label{f22}
\end{figure}

The linear response was checked against numerical simulations of Eq. \eqref{eq:motion}, by starting from an initial width $\sigma(t=0)=\sigma_0+\delta$, where $\delta$ is a small perturbation. In the linear regime ($\delta \ll \sigma_0$), the oscillation frequency is independent of $\delta$ and close to the analytical value obtained above (see Fig. \ref{s1}).
However, one of the interesting features of the variational approach is that it is not limited to the linear regime, and allows us to investigate the dependence of the frequency with the excitation strength even for relatively large values of $\delta$.
Figure \ref{f2} shows some values of $\Omega$ obtained from the numerical solution of Eq. (\ref{eq:motion}), for different values of $\delta$. The frequency starts to deviate from the linear result for $\delta \approx 0.1$, which is roughly $5\%$ of the ground state width $\sigma_0$. Nonlinear effects appear to again redshift the frequency.
For a strong excitation ($\delta=0.5$), the frequency spectrum $|\sigma(\omega)|$ is displayed in Fig. \ref{f22}, showing a principal peak just below 30~eV (in accordance with Fig. \ref{f2}) along with a few higher harmonics.

\begin{table}
  \centering
\begin{tabular}{|c|c|c|}
\hline
$\,L^2_{ext}$ (a.u.) &  $\,\sigma_0$ (a.u.)  & $\,\Omega$ (eV) \\
\hline
\hline
0 & 1.719 & 33.81\\
\hline
2.5 & 1.720 & 33.79\\
\hline
5 & 1.721 & 33.76\\
\hline
7.5 & 1.722 & 33.74\\
\hline
10 & 1.724 & 33.72\\
\hline
30 & 1.734  & 33.53\\
\hline
50 & 1.744 & 33.34 \\
\hline
100 & 1.768 & 32.81 \\
\hline
200 & 1.817 & 31.57 \\
\hline
\end{tabular}
\caption{Values of the ground-state width $\sigma_0$ and linear response frequency $\Omega$ for different values of the extrinsic angular momentum $L_{ext}$.}
\label{t2}
\end{table}

\subsection{Extrinsic angular momentum}
In all the above results, we always assumed that the pressure tensor is isotropic, which implies that the tangential part of the pressure cancels the centrifugal potential, as was discussed in Sec. \ref{sec:model}.
However, the external excitation of the monopole mode (either optical or using EELS) may also impart a global angular momentum $L_{ext}$ to the $\rm C_{60}$ molecule \cite{Schuler2016}. Here, we briefly investigate the relevance of this effect.

This extrinsic angular momentum is modeled by a centrifugal potential $V_{ext}=L^2_{ext}/(2r^2)$.
The corresponding term in the Lagrangian density can be computed using the Ansatz of the electron density, Eq. \eqref{eq:ansatz}, to obtain:
\begin{equation}
U_L(\sigma)=\frac{4\pi}{N}\int_0^\infty n\,V_{ext}\,r^2\,dr=\frac{L_{ext}^2}{30\,\sigma^2},
\end{equation}
which should be added to the potential $U(\sigma)$ appearing in Eq. \eqref{eq:motion}.

The results for the ground state width and linear frequency are summarized in Table \ref{t2}. As expected, $\sigma_0$ increases slightly with increasing angular momentum. However, there is no significant change in the linear frequency $\Omega$ compared to the case $L_{ext}=0$, even when the extrinsic angular momentum much exceeds the intrinsic value $\langle L^2\rangle=37.5$ mentioned in Sec. \ref{sec:model}.

\section{Numerical simulations}\label{sec:numerical}

\subsection{QHD model}

In this section, we perform numerical simulations of the monopole mode by directly solving the full nonlinear QHD equations (\ref{8a1})-(\ref{8a2}) and \eqref{eq:poissonfluid}, with spherical symmetry. The numerical methods relies on the property that these equations (for $n$ and $u$) can be rewritten in the form of an ancillary nonlinear Schr\"odinger equation for a pseudo-wavefunction defined as $\Psi \equiv \sqrt{n} e^{i\theta}$, with $u=\partial_r \theta$:
\begin{equation}
i \frac{\partial\Psi}{\partial t} = -\frac{1}{2}\Delta_r \Psi + W \Psi ,  \label{eq:pseudoschrod}
\end{equation}
where $W=-V_H+V_{X,C}+V_{ps} + \int^n {{P(n')}\over n'} dn'$.
The above Schr\"odinger equation is then solved numerically using a standard finite-difference Crank-Nicolson scheme, together with Poisson's equation to obtain the Hartree potential $V_H(r)$. The computational box is $r \in [0, R_{max}]$, with $R_{max} = 80 \gg R$ and boundary conditions: $\Psi(R_{max})= V_H(R_{max})=0$ and $\Psi'(0)= V_H'(0)=0$, where the apex here stands for differentiation with respect to $r$.

\begin{figure}
  \centering
  \includegraphics[width=0.45\paperwidth]{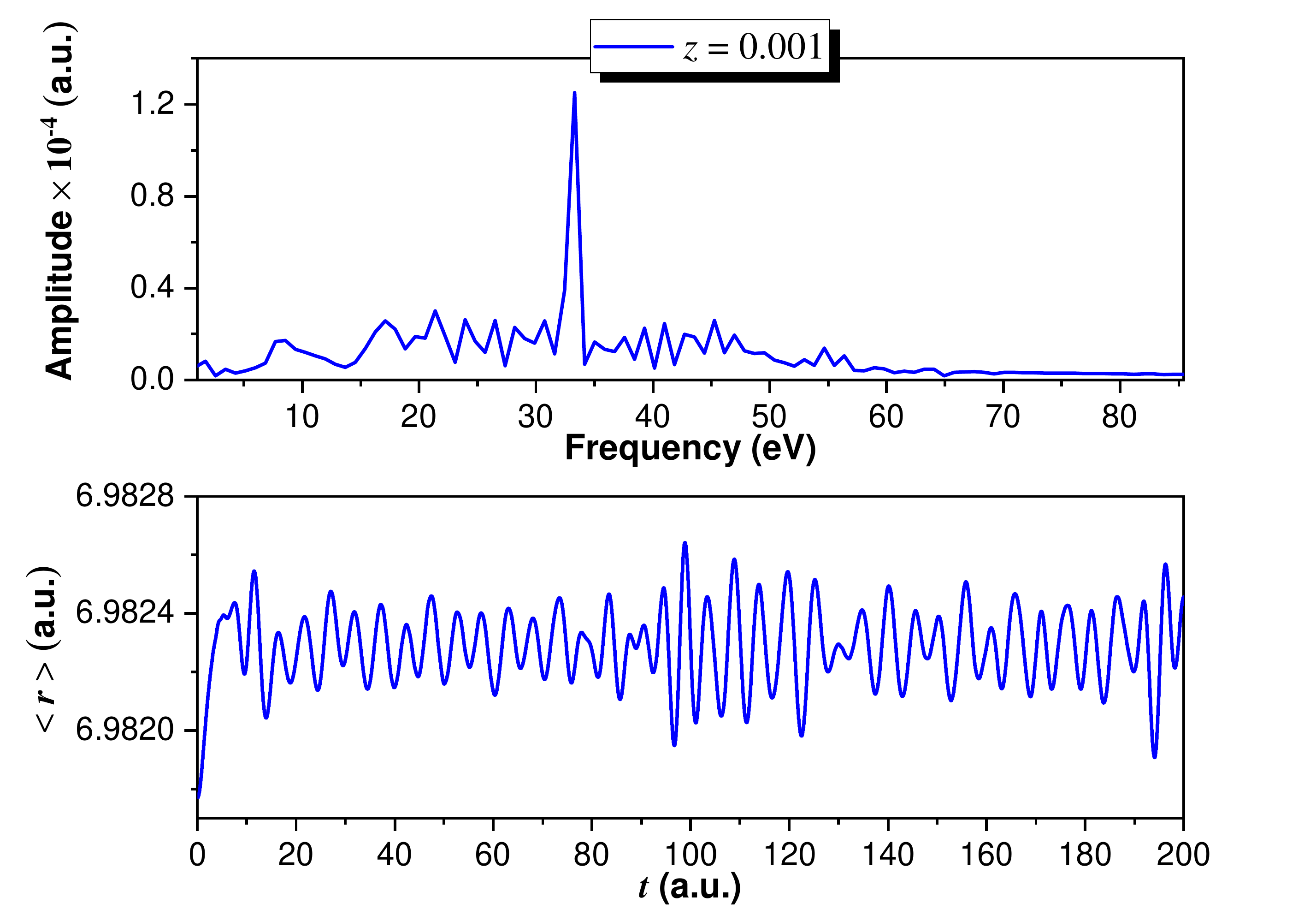}\\
  \caption{Evolution of the average radius of the electron cloud $\langle r \rangle(t)$ and corresponding frequency spectrum obtained from a direct solution of the dynamical QHD equations. The perturbation is an instantaneous Coulomb potential, as in Eq. \eqref{eq:excitation}, with $z=0.001$.} \label{fig:time-freq-z1}
\end{figure}
\begin{figure}
  \centering
  \includegraphics[width=0.45\paperwidth]{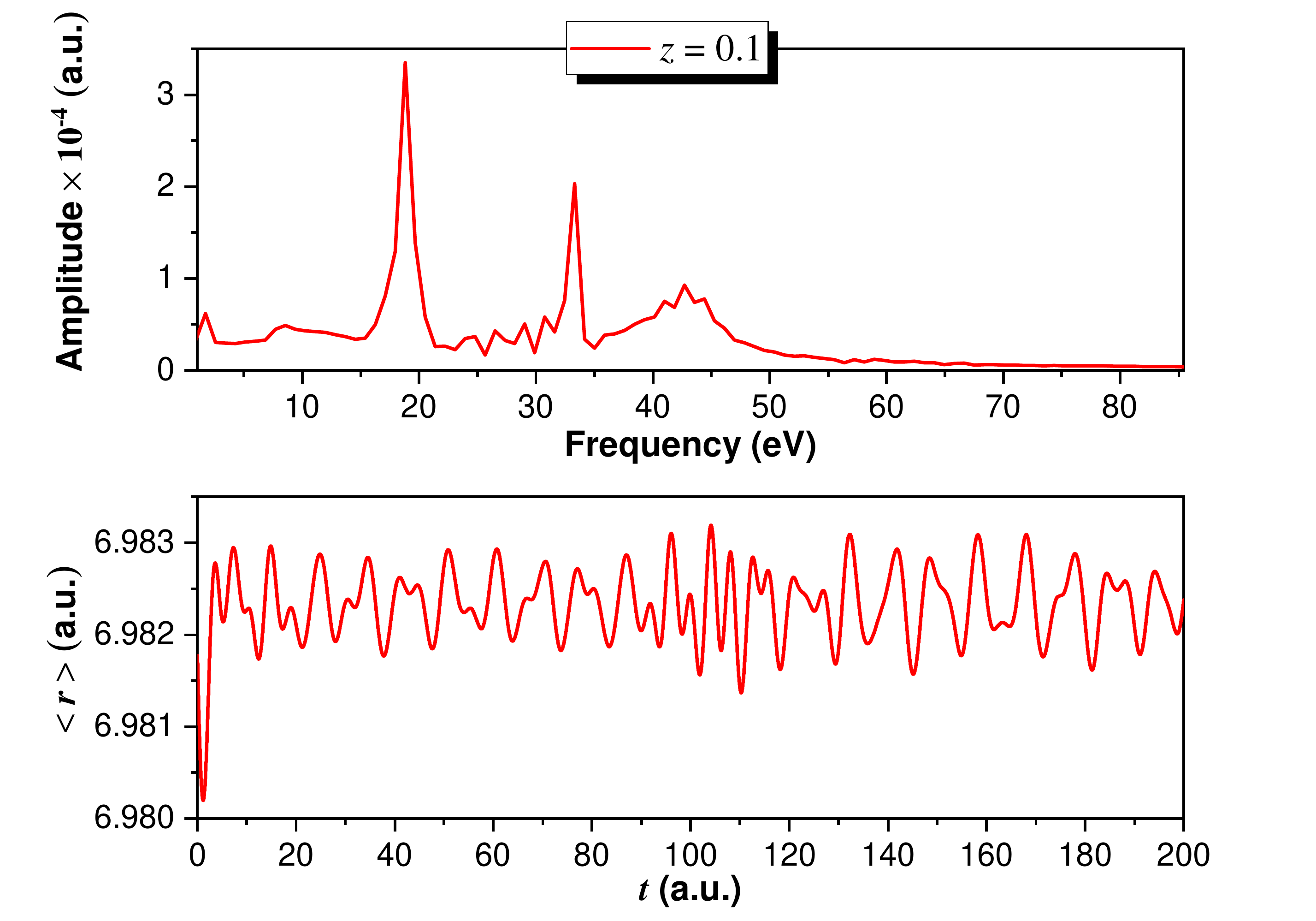}\\
  \caption{Same as Fig. \ref{fig:time-freq-z1}, with $z=0.1$.}\label{fig:time-freq-z2}
\end{figure}

First, we would like to verify the results obtained in the preceding sections on the linear response. For this, we need to compute the ground state of the system. This can be done by solving Eq. \eqref{eq:pseudoschrod} in the ``imaginary time" $\tau=i\,t$. This substitution transforms the above Schr\"odinger equation into a diffusion-like equation, which naturally relaxes to a steady-state solution that can be identified as the ground state of our system \cite{Crouseilles2008}. This method is used here to obtain the ground-state profiles of the electron density and the various potentials shown in Sec. \ref{sec:lagrangian}.

Next, the ground state must be slightly perturbed to induce some dynamical processes. As a possible excitation, we use an instantaneous Coulomb potential applied at $t=0$:
\begin{equation}
V_{ext}(r,t) = \frac{z}{r}\, \delta(t),
\label{eq:excitation}
\end{equation}
where $\delta$ is the Dirac delta function and the fictitious charge $z$ quantifies the magnitude of the perturbation.

In order to study the system response to such excitation, we analyze the time evolution of the average radius of the electron cloud:
\begin{equation}
\langle r \rangle = {1\over N}\, \int_0^{\infty} r\, n(r,t) \,4\pi r^2 dr.
\label{eq:raverage}
\end{equation}
The result of two simulations for two values of the excitation amplitude $z$ are shown in Figs. \ref{fig:time-freq-z1} and \ref{fig:time-freq-z2}, both in the time and in the frequency domains.
At very low amplitude ($z=10^{-3}$, Fig. \ref{fig:time-freq-z1}), we measure a monopole frequency $\Omega \approx 33.2 \,\rm eV$, very close to the semianalytical result obtained in Sec. \ref{sec:lagrangian}. This is the expected breathing mode, at a frequency close to the plasmon frequency, slightly redshifted because of the spill-out and other quantum effects.

We can actually show that this redshift is mainly due to the spill-out effect by resorting to a simple argument. It is expected that the spillout-corrected frequency reads as:
\begin{equation}
\Omega^2 = \omega_p^2 \, \left(1-{N_{out}\over N} \right)= \omega_p^2 \,{N_{in} \over N} \,,
\label{eq:spillout}
\end{equation}
where $N_{out}$ and $N_{in}$ are respectively the number of electrons outside and inside the ionic jellium, e.g.: $N_{in} = \int_{R_1}^{R_2} n_{gs}\, 4\pi^2 r^2\, dr$.
Using this simple prescription and the ground state density $n_{gs}$ obtained from the QHD code, we obtain $\Omega \approx 33.5\,\rm eV$, in very good agreement with both the variational semianalytical result and the QHD simulations. This reinforces our suspicion that the main correction to the classical Mie frequency comes from the nonlocal spill-out effect.

More surprisingly, a second mode appears at larger excitations (but still in the linear-response domain) and becomes dominant for $z=0.1$ (Fig. \ref{fig:time-freq-z2}). Its frequency is roughly 19~eV, which is intriguingly close to the surface plasmon frequency of a spherical nanoparticle $\omega_p/\sqrt{3}$, further redshifted as in Eq. \eqref{eq:spillout} because of the spill-out.

\begin{figure}
  \centering
  \includegraphics[width=0.45\paperwidth]{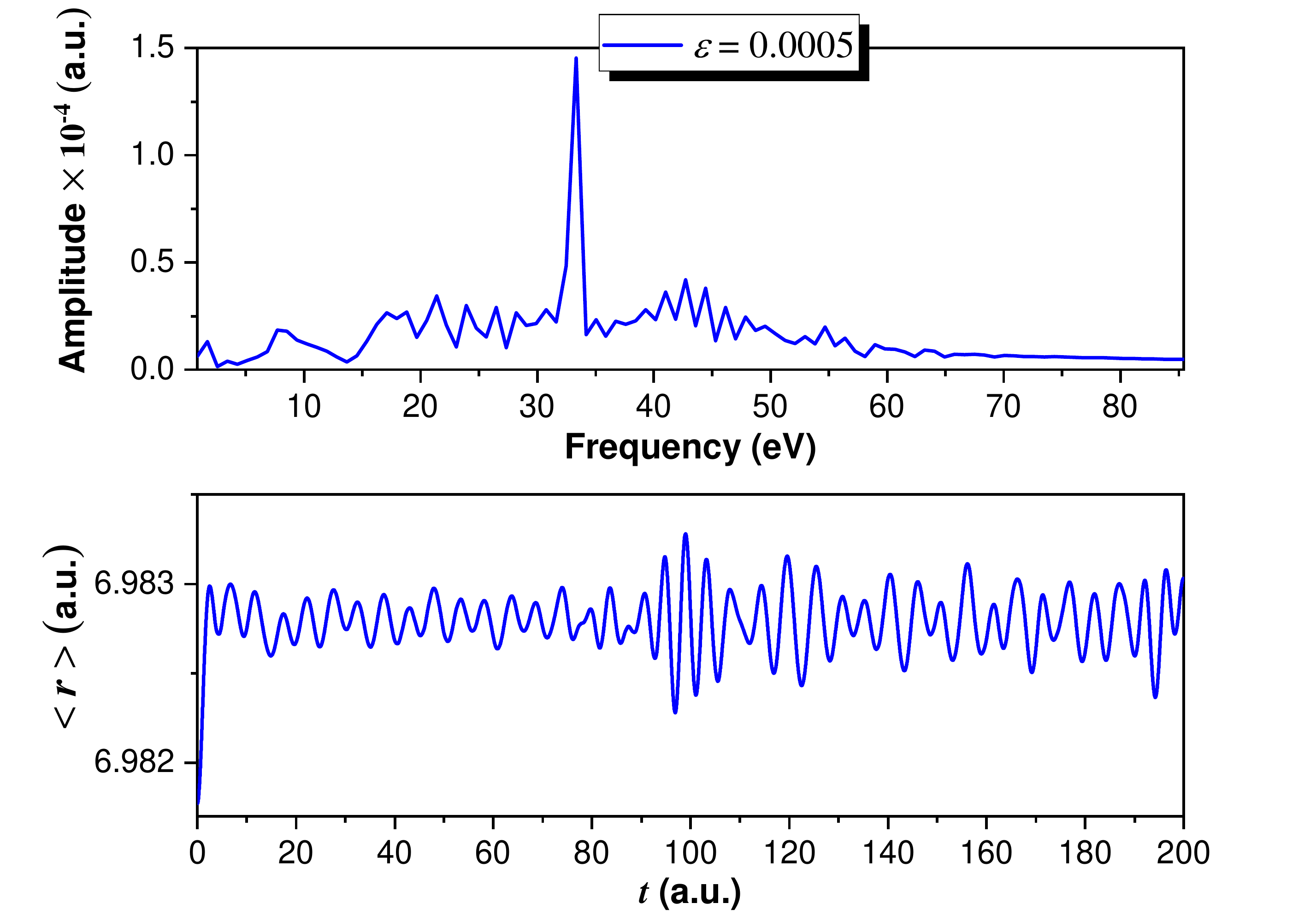}\\
  \caption{Evolution of the average radius of the electron cloud $\langle r \rangle(t)$ and corresponding frequency spectrum obtained from a direct solution of the dynamical QHD equations. The perturbation is a small shift of the ion background of a distance $\varepsilon=5 \times 10^{-4}$.}\label{fig:time-freq-d1}
\end{figure}

\begin{figure}
  \centering
  \includegraphics[width=0.45\paperwidth]{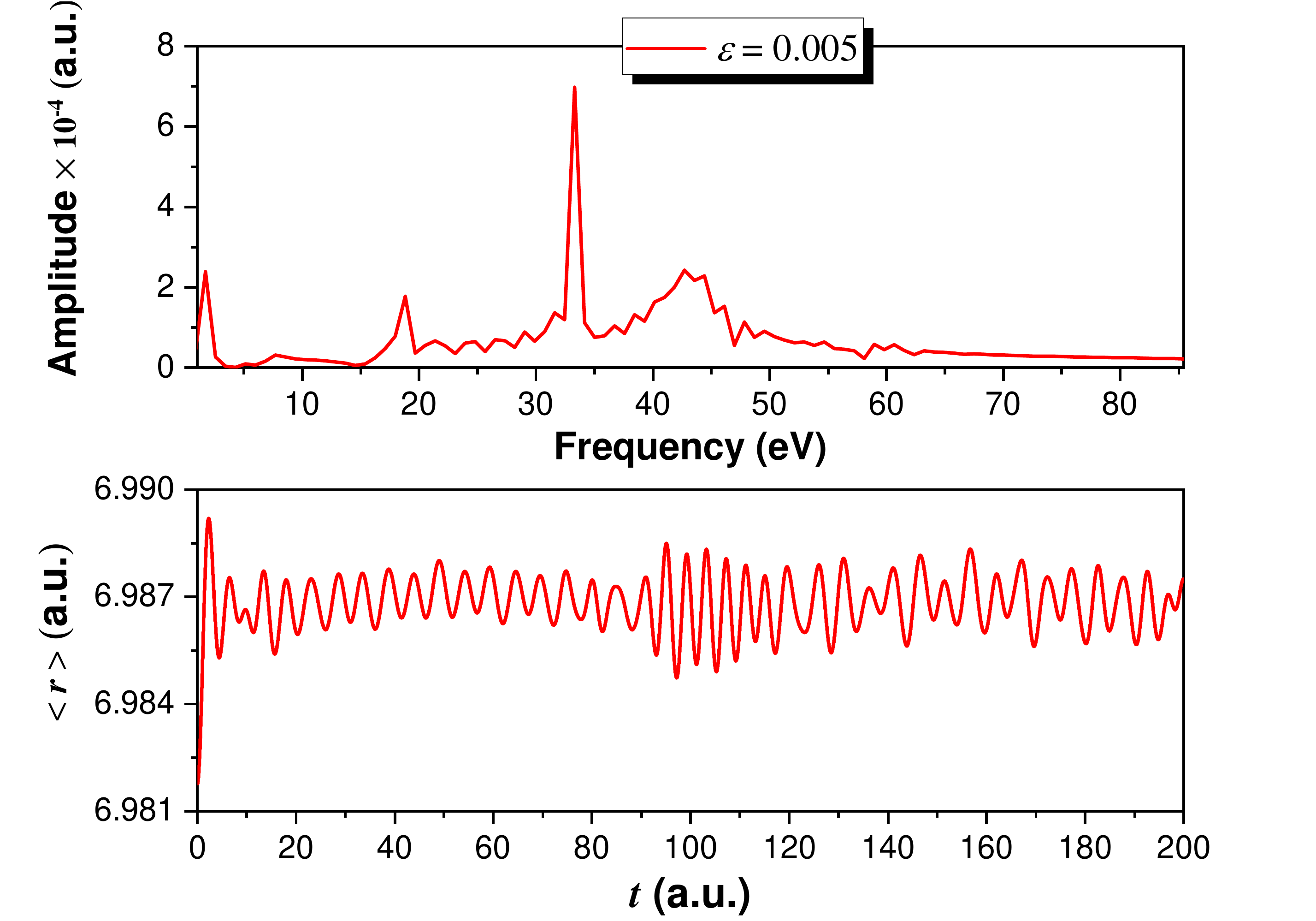}\\
  \caption{Same as Fig. \ref{fig:time-freq-d1} with $\varepsilon=5 \times 10^{-3}$.}\label{fig:time-freq-d2}
\end{figure}

\begin{figure}
  \centering
  \includegraphics[width=0.45\paperwidth,angle=0]{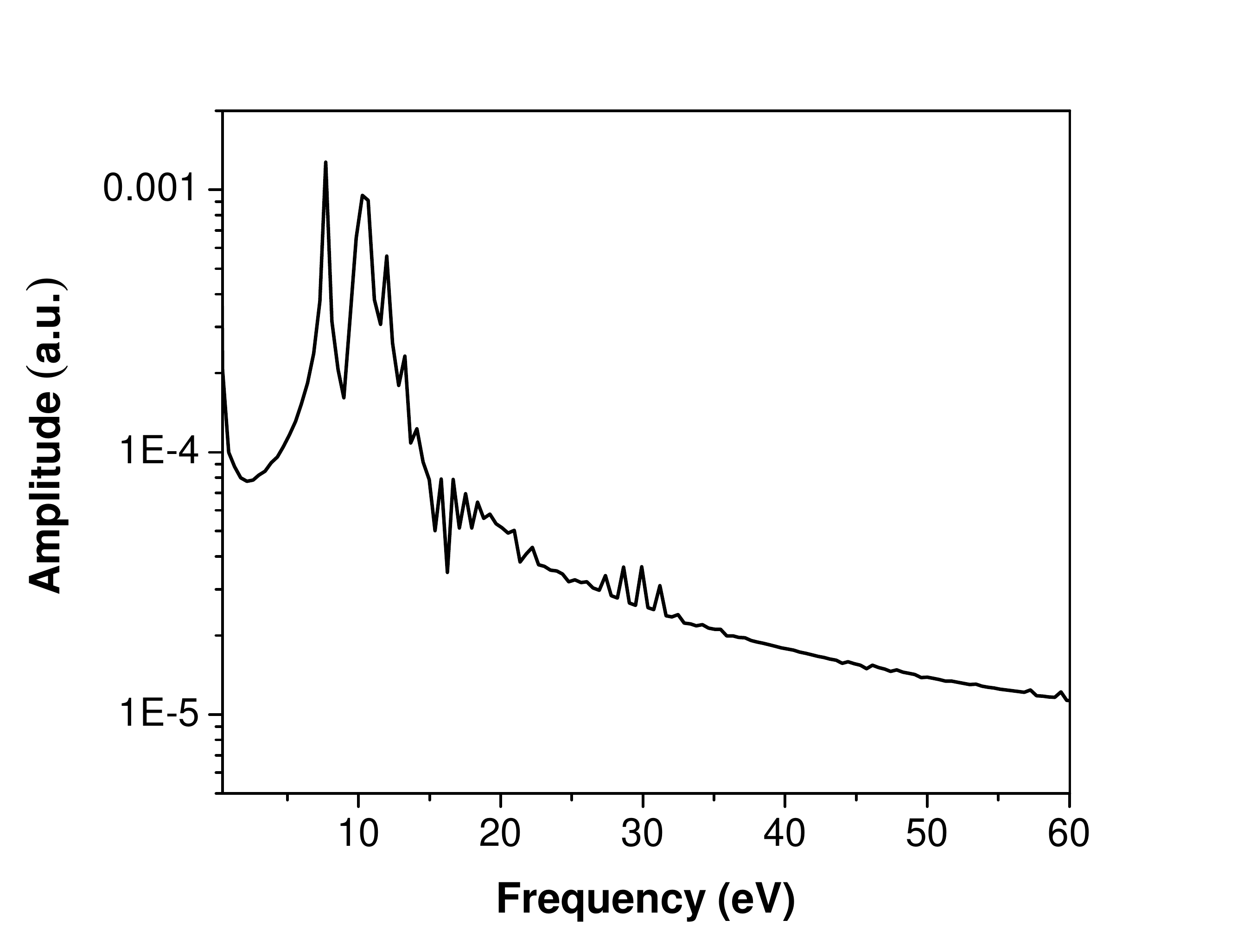}
  \caption{Free response spectrum obtained with the QHD model, for a Coulomb-like perturbation with $z=0.1$.}\label{fig:c60free}
\end{figure}

We interpret these two modes with the following arguments. For small values of $z$, the exciting force $F_{ext} = -V'_{ext}$ varies little across the electron density profile. This excitation thus simply shifts the electron cloud radially, and the latter starts to oscillate in a ``dipole-like" way around its equilibrium. This is the standard plasmonic breathing mode at a frequency close to $\omega_p$.
We also recall that this mode was obtained through the classical Mie theory for a spherical shell with internal and external radii $R_1$ and $R_2$, assuming a flat electron density inside the shell and thus neglecting the spill-out effect \cite{Lundqvist1979,Ostling1993}.
The Mie theory predicts the following frequencies for excitations of angular momentum $l$:
\begin{equation}
\Omega_\pm^2 = {{\omega_p^2} \over 2}\, \left(1 \pm \frac{1}{2l+1}\sqrt{1+4l(l+1) \eta^{2l+1}} \right)\,,
\label{eq:multipole}
\end{equation}
where $\eta=R_1/R_2$. There are therefore two frequencies for each multipolar mode of angular momentum $l$, except for the monopole case ($l=0$), for which only $\Omega_+=\omega_p$ is meaningful, whereas $\Omega_{-}=0$.
For such a spherical dipole-like mode, the induced charged density is localized around the inner and outer radii of the shell, which is compatible with an excitation that does not vary much within the electron cloud, in accordance with our case at low $z$.

In contrast, for larger perturbations, the gradient of the external force $F_{ext}$ becomes noticeable and induces another monopolar modes that cannot be accounted for by the Mie theory. This effects is enhanced by the large spill-out present in the case of $\rm C_{60}$, for which the electron density is far from the homogeneous profile that is assumed in the standard Mie theory. The fact that this extra monopolar mode has a frequency close to $19\,\rm eV$ is not yet explained and may be due to the specific profile of the electron density in the $\rm C_{60}$ molecule.

In order to check the above hypotheses on the origin of the $19\,\rm eV$ mode, we repeated the analysis using a different perturbation, more similar to the one generally assumed to obtain the result of Eq. \eqref{eq:multipole}. To do so, after computing the ground state, we shift radially the ion background jellium by a very small amount $\varepsilon \ll R$, and then let the electron gas evolve self-consistently. This type of perturbation is indeed localized at the ionic jellium boundaries.
The results are shown in Figs. \ref{fig:time-freq-d1} and \ref{fig:time-freq-d2} for $\varepsilon=5\times 10^{-4}$ and $\varepsilon=5\times 10^{-3}$ respectively, which induce center-of-mass oscillations of the same order of magnitude as the Coulomb-type excitation described earlier.
The verdict is rather clear: in this case, the monopolar plasmon mode at 33~eV is always largely dominant, in accordance with the standard spillout-corrected Mie theory (a small peak around 19~eV is nevertheless visible in the higher excitation case).

Finally, we performed a simulation for the so-called free response of the system, for which the effective (total) potential is kept fixed and equal to that of the ground state. Doing so effectively cuts all the electron-electron interactions, so that the response is reduced to the single-particle excitations, and collective self-consistent modes are suppressed. The spectrum, obtained for a Coulomb-like perturbation with $z=0.1$, is shown in Fig. \ref{fig:c60free}. As expected, the two modes at 19~eV and 33~eV, observed in the fully self-consistent simulations of Figs. \ref{fig:time-freq-z1} and \ref{fig:time-freq-z2}, do not appear in the free-response spectrum. This result constitutes further confirmation that these are indeed collective many-electron modes.

The above set of simulations lead us to conclude that: (i) when the excitation is a spatially homogeneous kick and localized at the system's boundaries, the response is the one predicted by the spillout-corrected Mie theory for the same configuration; (ii) when the excitation is spatially modulated through the electron density, a second peak appears at lower energy.

\subsection{TDDFT calculations}

\begin{figure}
  \centering
  \includegraphics[width=0.4\paperwidth]{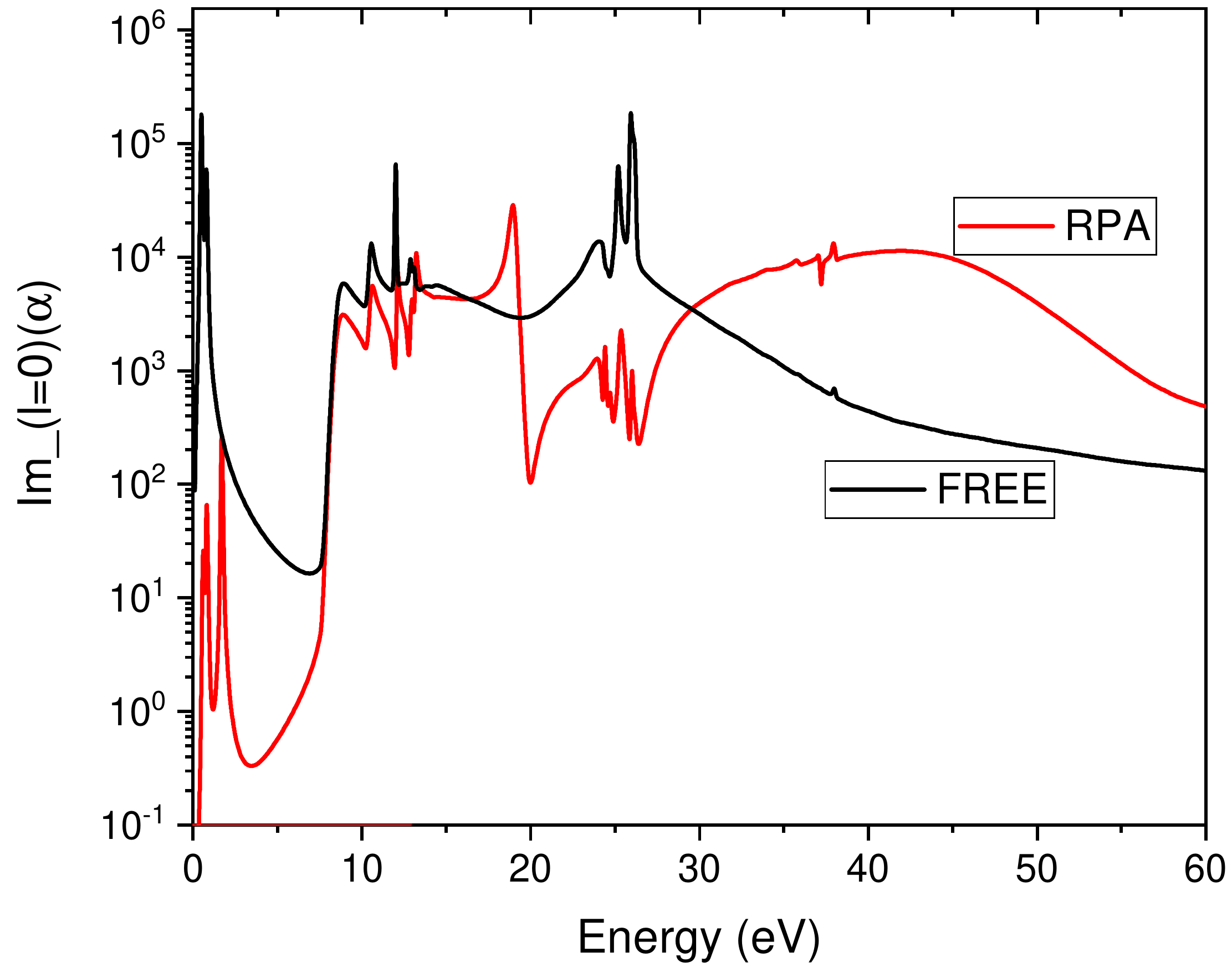}\\
  \caption{Imaginary part of the monopolar polarizability $\alpha$ as a function of the excitation energy, for the correlated response (RPA, red curve) and for the free response (black curve).}\label{fig:Im_alpha}
\end{figure}

\begin{figure}
  \centering
  \includegraphics[width=0.4\paperwidth]{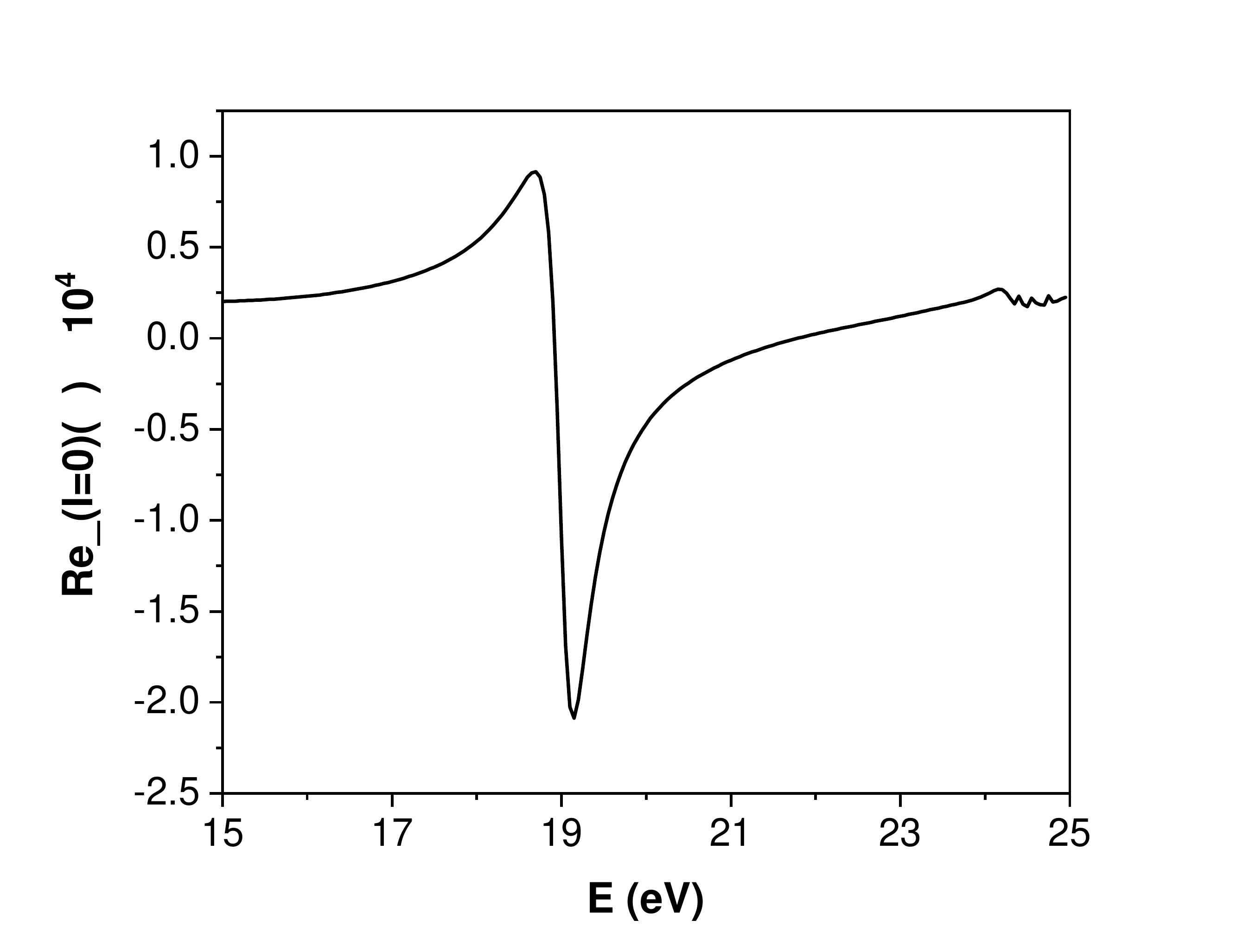}
    \includegraphics[width=0.4\paperwidth]{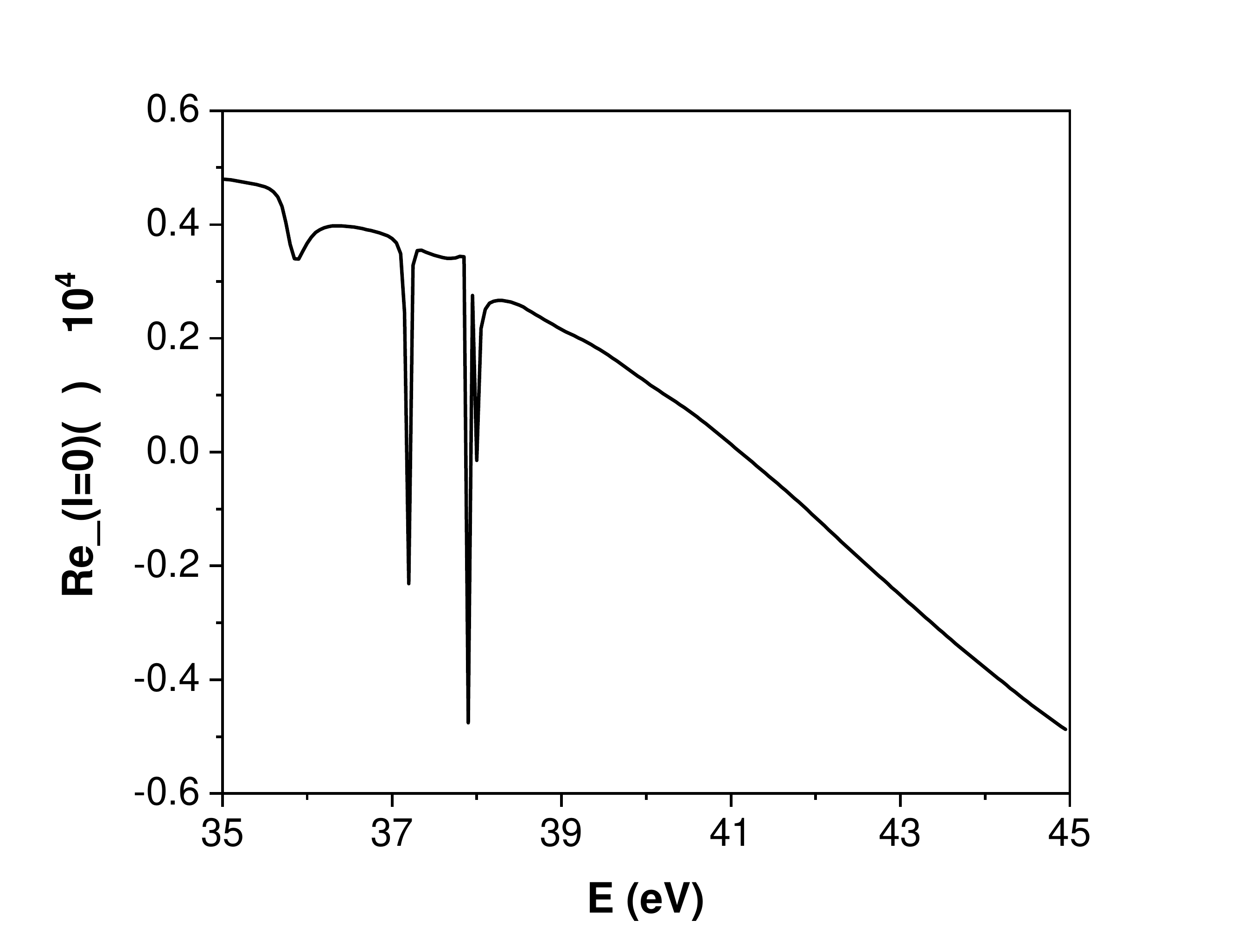}
  \caption{Two zooms of the real part of the monopolar polarizability $\alpha$, in the ranges 15-25~eV (left panel) and 35-45~eV (right panel).}\label{fig:Re_alpha}
\end{figure}

In order to better understand the nature of the observed monopolar modes, we also performed some simulations based on a TDDFT (Kohn-Sham) approach in the linear response regime \cite{Maurat2009}.
{\revfirst These simulations use the same jellium model, pseudopotential and exchange-correlation functionals as the corresponding QHD runs (see also Appendix \ref{A1} for more details)}.

{\revsec
The spatial form of the external potential $\delta V$ is chosen to be a pure multipole mode. In order to excite only the spherically symmetric modes $l=0$ (but not the dipole or other multipolar modes), one must take $\delta V \propto r^2\, Y_{00}({\mathbf r}/r)$, where $Y_{00} \sim \rm const.$ is the $l=m=0$ spherical harmonic. This corresponds to a physical situation where the momentum transfer vanishes, which is precisely the regime investigated with the QHD approach.
Other types of excitations, not explored here, may also be of interest. For instance, a plane-wave field may be used to model electron energy loss scattering as was done in some recent works \cite{Schuler2015}.
}

The imaginary part of the monopolar polarizability $\alpha$ is represented in Fig. \ref{fig:Im_alpha}, for both the correlated (RPA) and the free response. In the correlated case, there is a very broad peak extending from 35 to 45~eV, which we can attribute to the plasmonic monopolar mode. The broadness and blue shift with respect to the plasmon frequency may be attributed to the fact that, at these high energies, the coupling with the continuous part of the spectrum is rather significant (the ionization potential is about 7.5~eV). A similar broadening was also observed for the corresponding surface plasmon dipolar mode \cite{Madjet2008}.
A second peak appears near 19~eV, which is reminiscent of the peak we observed at the same energy in the QHD simulations.
These two peaks are absent from the free response spectrum, clearly suggesting that they represent collective modes. Other peaks, presumably due to single particle excitations, are common to the two spectra.

In order to better understand the character of these two peaks, we plot the real part of the monopolar polarizability (Fig. \ref{fig:Re_alpha}), zoomed in on the relevant energy scales. For collective modes, one would expect that the real part goes through zero at resonance. This is definitely the case for the lower energy mode, for which $\rm Re\, \alpha$ changes sign very abruptly near 19~eV.
For the higher energy peak, some smoother sign reversal is observed near 41~eV, which allows us to locate more precisely the resonant frequency. Taking into account the blue shift mentioned above, this is not too far from the 33.8~eV predicted by the QHD theory.

All in all, both the QHD and TDDFT approaches predict the existence of two collective monopolar modes, a sharper one at lower energy (19~eV) and a much broader one in the range 30-40~eV.
We also mention that two similar monopolar volume modes were observed in atomistic ab-initio TDDFT simulations of EELS in fullerenes \cite{Schuler2015}. The measured frequencies were around 24~eV and 42~eV (with rather extended peaks), which is in broad agreement with our results.
{\revsec
As the authors point out, the low-frequency mode has a quantum origin that may be attributed to the electron spill-out -- this is a further sign that the QHD method is capable to deal with those subtle effects.
From closer inspection of the Fig. 5 in the above work, it is clear that the volume plasmon contribution to the spectrum increases for decreasing scattering angle (i.e., for decreasing momentum transfer $q$) and becomes presumably maximal for $q=0$, which was the value used for all our simulations.

Finally, collective multipolar excitations in the EELS spectra of $\rm C_{60}$ molecules were also described in a recent theoretical/experimental study \cite{Bolognesi2012}, with the measured spectra agreeing well with the simulations of Sch\"uler {\it et al.} \cite{Schuler2015} (see their Fig. 5d-f).
The monopolar volume mode was not detectable in the experiments, presumably because it remains somewhat small compared to the surface modes, even at relatively low momentum transfer. However, as discussed in the preceding paragraph, this may change when $q \to 0$.
}

\section{Conclusions}
The $\rm C_{60}$ molecule has been to object of intense investigations since its experimental discovery in 1985. Conceptually, it lies at the border between large molecular systems and small nano-objects, and shares with the latter many dynamical properties. In particular, it exhibits typical plasmon collective resonances, which have been studied in depth for the case of a dipolar excitation.

{\revsec In the present work, our purpose was twofold}. Firstly, we provided an illustration of how quantum hydrodynamic methods can be successfully applied to many-electron systems like $\rm C_{60}$, for which a detailed ab-initio description would constitute a {\revsec far more complex} computational problem. The QHD approach provides a rather good approximation of the ground state profiles, both for the electron density and the effective potential.
The QHD equations can be further reduced, through an appropriate Ansatz, to a simple macroscopic equation describing the evolution of the width of the electron density. This equation provides analytically the linear response frequency of the system, and can be easily solved numerically to explore the nonlinear regime.

Our second purpose was to characterize plasmonic breathing modes (monopolar electronic modes with $l=0$) for the $\rm C_{60}$ molecule, which have been much less studied than the corresponding dipolar modes ($l=1$). Although more difficult to excite and detect experimentally, monopolar modes can nowadays be driven using electron energy loss spectroscopy (EELS).

We used the three approaches described above to tackle this problem: (i) the analytical QHD/variational method, (ii) the direct numerical solution of the full QHD equations, and (iii) an ab-initio TDDFT approach.
The analytical approach revealed one collective resonance at 33~eV, near the bulk plasmon frequency but redshifted mainly because of the spill-out effect. This is the standard monopole resonance predicted by Mie theory \cite{}, corresponding to a perturbed density localized at the inner and outer radii of the system.
The numerical QHD and TDDFT approaches pointed at a second collective resonance at lower energy (19~eV). We speculated that this second resonance corresponds to bulk modulations of the electron density.
The theoretical characterization of these collective resonances may hopefully pave the way to their experimental observation.

\section*{Acknowledgements}
This project has received funding from the European Union's Horizon 2020 research and innovation programme under grant agreement No 701599 -- QHYDRO -- H2020-MSCA-IF-2015.

\appendix

\section{Ground-state configuration of $\rm C_{60}$}\label{A1}

{\revsec
As described in full details in our earlier work \cite{Maurat2009}, the ionic background of $\rm C_{60}$ has been treated in the spherical jellium approximation following the model developed by Bauer \textit{et al.} \cite{Bauer2001}. In the latter, the charge of the real ionic cores is replaced by a constant positive background uniformly distributed over a spherical shell.

Moreover, in order to ensure two important features resulting from quantum chemical calculations \cite{Troullier1992} -- namely that (i) the two highest occupied molecular orbitals are of $1h$ and $1g$ character and (ii) the HOMO level is approximately half-filled -- we have employed the procedure developed by Madjet \textit{et al.} \cite{Madjet2008}. This procedure leads to a partial filling of the spherical orbitals. The obtained quantum numbers, occupation numbers, and energies are listed in Table \ref{tab:config}. In the past, this model was successfully employed for the modelling of various physical processes requiring the knowledge of the electronic properties of $\rm C_{60}$ \cite{Cormier2003,Rentenier2008}.
}

\begin{table*}[ht!]
\begin{tabular}{|c|c|c|c|}
  \hline
  $p_l$ & $l$ & $n_r$ & $E$ (Hartree) \\ \hline
  2  &   0 &   0   &    $-1.39$ \\
  6  &   1 &   0   &    $-1.37$ \\
  10 &   2 &   0   &    $-1.32$ \\
  14 &   3 &   0   &    $-1.25$ \\
  18 &   4 &   0   &    $-1.16$ \\
  22 &   5 &   0   &    $-1.05$ \\
  26 &   6 &   0   &    $-0.914$ \\
  30 &   7 &   0   &    $-0.763$ \\
   2 &   0 &   1   &    $-0.639$ \\
   6 &   1 &   1   &    $-0.611$ \\
  34 &   8 &   0   &    $-0.594$ \\
  10 &   2 &   1   &    $-0.557$ \\
  14 &   3 &   1   &    $-0.481$ \\
  18 &   9 &   0   &    $-0.409$ \\
  18 &   4 &   1   &    $-0.387$ \\
  10 &   5 &   1   &    $-0.276$ \\
  \hline
\end{tabular}
\caption{Quantum ground-state structure of the $\rm C_{60}$ molecule as used in the derivation of the QHD model. The columns represent, from left to right: the occupation numbers $p_l=2(2l+1)$, azimuthal quantum number $l$, radial quantum number $n_r$ ($n_r=0$ for $\sigma$ electrons and $n_r=1$ for $\pi$ electrons), and energy $E$.}
\label{tab:config}
\end{table*}

\section{Detailed calculations for the variational approach}\label{A2}

The fluid set of equations can be exactly represented by a Lagrangian density $\mathcal{L}(n,\theta,V_H)$, where the function $\theta$ is related to the mean velocity through $u=\partial\theta/\partial r$. The expression for the Lagrangian density is the following:
\begin{eqnarray}
&&\mathcal{L}=n\left[\frac{1}{2}\left(\frac{\partial\theta}{\partial r}\right)^2+\frac{\partial\theta}{\partial t}\right]+\frac{1}{8n}\left(\frac{\partial n}{\partial r}\right)^2+\frac{3}{10}(3\pi^2)^{2/3}n^{5/3} -\frac{3}{4\pi}(3\pi^2)^{1/3}n^{4/3}\nonumber\\
&&\,\,\,\,\,\,\,\,+nV_{ps}-(n-n_i)V_H-\frac{1}{8\pi}\left(\frac{\partial V_H}{\partial r}\right)^2.\label{12}\
\end{eqnarray}
By taking the Euler-Lagrange equations with respect to the fields $n$, $\theta$ and $V_H$, one recovers exactly the fluid set of equations.

Now, the Lagrangian function can be defined as:
\begin{equation}
L(\sigma,\dot{\sigma})=\frac{1}{N}\int \mathcal{L}\,d\mathbf{r}=\frac{4\pi}{N}\int \mathcal{L}\,r^2dr. \label{eq:lagrangian}
\end{equation}
Substituting Eq. (\ref{12}) into Eq. \eqref{eq:lagrangian} and using the definition of $V_{ps}$, we get
\begin{eqnarray}
&& L(\sigma,\dot{\sigma})=\frac{4\pi}{N}\Biggl[\overbrace{\frac{1}{2}\int_0^\infty n\left(\frac{\partial\theta}{\partial r}\right)^2r^2dr}^{I_1}+\overbrace{\int_0^\infty n\frac{\partial \theta}{\partial t}\,r^2 dr}^{I_2}+\overbrace{\frac{1}{8}\int_0^\infty\frac{1}{n}\left(\frac{\partial n}{\partial r}\right)^2r^2dr}^{I_3} \nonumber\\
&&+\overbrace{\frac{3}{10}(3\pi^2)^{2/3}\int_0^\infty n^{5/3}r^2dr}^{I_4}-\overbrace{\frac{3}{4\pi}(3\pi^2)^{1/3}\int_0^\infty n^{4/3}r^2dr}^{I_5}-\overbrace{V_0 \int_{R_1}^{R_2}n r^2\,dr}^{I_6} \nonumber\\
&&- \overbrace{\int_0^\infty \left\{(n-n_i)V_H+\frac{1}{8\pi}\left(\frac{\partial V_H}{\partial r}\right)^2\right\}r^2\,dr}^{I_7}\,\Biggr].\label{15}\
\end{eqnarray}
Using the expressions of $n$ and $\theta$ provided in the main text, the first six integrals can be calculated as follows:
\begin{eqnarray}
&&I_1=\frac{4\pi}{N}\frac{1}{2}\int_0^\infty n\left(\frac{\partial\theta}{\partial r}\right)^2r^2dr=\frac{17\dot{\sigma}^2}{2},\nonumber\\
&&I_2=\frac{4\pi}{N}\int_0^\infty n\frac{\partial \theta}{\partial t}\,r^2dr=\frac{4\pi}{N}\frac{1}{2}\left(\frac{\ddot{\sigma}}{\sigma}-\frac{\dot{\sigma}^2}{\sigma^2}\right)\int_0^\infty n\,r^4\,dr = \frac{17}{2}\left(\ddot{\sigma}\sigma-\dot{\sigma}^2\right)=\frac{17}{2}\left[\frac{d}{dt}(\dot{\sigma}\sigma)-2\dot{\sigma}^2\right]=-17\dot{\sigma}^2,\nonumber\\
&&I_3=\frac{4\pi}{N}\frac{1}{8}\int_0^\infty\frac{1}{n}\left(\frac{\partial n}{\partial r}\right)^2r^2dr=\frac{4\pi}{8\,N}\int_0^\infty n\left(\frac{k}{r}-\frac{r}{\sigma^2}\right)^2r^2dr=\frac{\alpha_1}{\sigma^2},  \nonumber\
\end{eqnarray}
\begin{eqnarray}
&&I_4=\frac{4\pi}{N}\frac{3}{10}(3\pi^2)^{2/3}\int_0^\infty n^{5/3}r^2dr=\frac{\alpha_2 N^{2/3}}{\sigma^2},\nonumber\\
&&I_5=-\frac{4\pi}{N}\frac{3}{4\pi}(3\pi^2)^{1/3}\int_0^\infty n^{4/3}r^2dr=-\left(\frac{3}{2}\right)^{5/6}\frac{9}{32\pi^{4/3}}\Gamma{\left(\frac{17}{6}\right)}\frac{N^{1/3}}{\sigma}=-\frac{\alpha_3N^{1/3}}{\sigma},\nonumber\\
&&I_6=-\frac{4\pi V_0}{N} \int_{R_1}^{R_2}n r^2\,dr=\frac{V_0}{\sigma^{15}}\left[-R_1\exp\left(-\frac{R_1^2}{2\sigma^2}\right)F_1(R_1,\sigma)+R_2\exp\left(-\frac{R_2^2}{2\sigma^2}\right)F_2(R_2,\sigma)\right]\nonumber\\
&&\,\,\,\,\,\,\,\,\,+V_0\left[\erf\left(\frac{R_1}{\sqrt{2}\sigma}\right)-\erf\left(\frac{R_2}{\sqrt{2}\sigma}\right)\right],  \nonumber\
\end{eqnarray}
where $\alpha_1\approx 0.258$, $\alpha_2\approx 0.045$, $\alpha_3\approx 0.091$, and the functions $F_1$ and $F_2$ are defined by
\begin{eqnarray}
&&F_1(R_1,\sigma)=R_1^{14}+15R_1^{12}\,\sigma^{2}+195R_1^{10}\,\sigma^{4}+2145R_1^{8}\,\sigma^{6}+19305R_1^{6}\,\sigma^{8}+135135R_1^{4}\,\sigma^{10}\nonumber\\
&&+675675R_1^{2}\,\sigma^{12}+2027025\sigma^{14},\\
&&F_2(R_2,\sigma)=R_2^{14}+15R_2^{12}\,\sigma^{2}+195R_2^{10}\,\sigma^{4}+2145R_2^{8}\,\sigma^{6}+19305R_2^{6}\,\sigma^{8}+135135R_2^{4}\,\sigma^{10}\nonumber\\
&&+375375R_2^{2}\,\sigma^{12}+2027025\sigma^{14}.
\end{eqnarray}

To perform the last integral, we use Poisson's equation and write $I_7$ in the following way:
{\small $$I_7=-\frac{1}{N}\int \left\{(n-n_i)V_H+\frac{1}{8\pi}\left(\frac{\partial V_H}{\partial r}\right)^2\right\}\,d\mathbf{r} = -\frac{1}{4\pi N}\int\nabla_r\cdot(V_H\nabla_r V_H)d\mathbf{r}+\frac{1}{8\pi N}\int\left(\frac{\partial V_H}{\partial r}\right)^2\,d\mathbf{r}.$$} The first (divergence) term disappears upon integration over space for reasonable boundary conditions, so that only the second integral is required. For evaluating the second integral, we decompose the Hartree potential as $V_H = V_i+V_e$, where $V_{i,e}$ are the contributions due to the ions and the electrons respectively, which satisfy the equations
\begin{eqnarray}
&&\Delta_r V_i=-4\pi n_i,\label{17a}\\
&&\Delta_r V_e=4\pi n, \label{17b}\
\end{eqnarray}
Thus the integral can be rewritten as
{\small\begin{equation}\label{17c}
I_7=\frac{1}{8\pi N}\int\left(\frac{\partial V_H}{\partial r}\right)^2\,d\mathbf{r}=\frac{1}{8\pi N}\left[\int\left(\frac{\partial V_i}{\partial r}\right)^2\,d\mathbf{r}+\int\left(\frac{\partial V_e}{\partial r}\right)^2\,d\mathbf{r}+2\int\left(\frac{\partial V_i}{\partial r}\right)\left(\frac{\partial V_e}{\partial r}\right)\,d\mathbf{r}\right].
\end{equation}}
By injecting the definitions of $n(r,t)$ and $n_i(r)$ into Eqs. (\ref{17a}) and (\ref{17b}) and integrating once, we compute the gradients $\partial V_i/\partial r$ and $\partial V_e/\partial r$ as
\begin{eqnarray}
&&\frac{\partial V_e}{\partial r}=\frac{4\pi \,A}{\sigma^{15}}\frac{1}{r^2}\Biggl[-r\,\exp\left(-\frac{r^2}{2\sigma^2}\right)G(r,\sigma)+2027025\,\sqrt{\frac{\pi}{2}}\,\sigma^{15}\erf\left(\frac{r}{\sqrt{2}\sigma}\right)\Biggr],\label{10a}\\
&&\frac{\partial V_i}{\partial r}=\frac{4\pi n_0}{3}\frac{1}{r^2}\Biggl[-\left(r^3-R_1^3\right)\mathcal{H}\left(r-R_1\right)+\left(r^3-R_2^3\right)\mathcal{H}\left(r-R_2\right)\Biggr],\label{10b}\
\end{eqnarray}
where $$G(r,\sigma)=r^{14}+15\,r^{12}\,\sigma^{2}+195\,r^{10}\,\sigma^{4}+2145\,r^{8}\,\sigma^{6}+19305\,r^{6}\,\sigma^{8}+135135\,r^{4}\,\sigma^{10}\\+675675\,r^{2}\,\sigma^{12}+2027025\,\sigma^{14}.$$
Now, the first integral of Eq. (\ref{17c}) does not contribute to the equations of motion because it does not depend on the dynamical variable $\sigma$. Let us evaluate the other two
integrals separately by using Eqs. (\ref{10a}) and (\ref{10b}), to get
\begin{equation}
\frac{1}{8\pi N}\int\left(\frac{\partial V_e}{\partial r}\right)^2\,d\mathbf{r}=\frac{1}{2N}\int_0^\infty\left(\frac{\partial V_e}{\partial r}\right)^2\,r^2\,dr=\frac{\alpha_4 N}{\sigma},\label{19a}
\end{equation}
and
\begin{eqnarray}
&&\frac{2}{8\pi N}\int\left(\frac{\partial V_i}{\partial r}\right)\left(\frac{\partial V_e}{\partial r}\right)\,d\mathbf{r}=\frac{1}{N}\int_0^\infty\left(\frac{\partial V_i}{\partial r}\right)\left(\frac{\partial V_e}{\partial r}\right)\,r^2\,dr\nonumber\\
&& = \frac{2\sqrt{2\pi}\,n_0}{2027025\,\sigma_{11}}\Biggl[R_1\exp\left(-\frac{R_1^2}{2\sigma^2}\right)K_1(R_1,R_2,\sigma)-R_2\exp\left(-\frac{R_2^2}{2\sigma^2}\right)K_2(R_1,R_2,\sigma)\Biggr]\nonumber\\
&&-\frac{2\pi n_0}{3}\Biggl[\erf\left(\frac{R_1}{\sqrt{2\sigma}}\right)K_3(R_1,R_2,\sigma)-\erf\left(\frac{R_2}{\sqrt{2\sigma}}\right)K_4(R_1,R_2,\sigma)\Biggr],\label{19b}\
\end{eqnarray}
where $\alpha_4\approx 0.114$, and the functions
\begin{eqnarray}
&&K_1(R_1,R_2,\sigma)=2R_1^{12}\,\sigma^{2}+80R_1^{10}\,\sigma^{2}+1810R_1^{8}\,\sigma^{4}+28020R_1^{6}\,\sigma^{6}+305130R_1^{4}\,\sigma^{8}\nonumber\\
&&+2231880R_1^{2}\,\sigma^{10}+675675(16-R_1+R_2)\,\sigma^{12},\\
&&K_2(R_1,R_2,\sigma)=2R_2^{12}\,\sigma^{2}+80R_2^{10}\,\sigma^{2}+1810R_2^{8}\,\sigma^{4}+28020R_2^{6}\,\sigma^{6}+305130R_2^{4}\,\sigma^{8}\nonumber\\
&&+2231880R_2^{2}\,\sigma^{10}+675675(16-R_1+R_2)\,\sigma^{12},\\
&&K_3(R_1,R_2,\sigma)=R_1^2(-2+R_1-R_2)+(16-R_1+R_2)\sigma^2,\\
&&K_4(R_1,R_2,\sigma)=R_2^2(-2+R_1-R_2)+(16-R_1+R_2)\sigma^2.\
\end{eqnarray}
Therefore the last integral becomes
\begin{eqnarray}\label{18a}\
&&I_7=\frac{\alpha_4 N}{\sigma}+\frac{2\sqrt{2\pi}\,n_0}{2027025\,\sigma_{11}}\Biggl[R_1\exp\left(-\frac{R_1^2}{2\sigma^2}\right)K_1(R_1,R_2,\sigma)-R_2\exp\left(-\frac{R_2^2}{2\sigma^2}\right)K_2(R_1,R_2,\sigma)\Biggr]\nonumber\\
&&-\frac{2\pi n_0}{3}\Biggl[\erf\left(\frac{R_1}{\sqrt{2\sigma}}\right)K_3(R_1,R_2,\sigma)-\erf\left(\frac{R_2}{\sqrt{2\sigma}}\right)K_4(R_1,R_2,\sigma)\Biggr].
\end{eqnarray}
Combining all the integrals, we can write the Lagrangian
\begin{equation}\label{18b}\
L(\sigma,\dot{\sigma})=-\frac{17\dot{\sigma}^2}{2}+U(\sigma),
\end{equation}
where
{\small\begin{eqnarray}\label{18c}\
&&U(\sigma)=\frac{\alpha_1}{\sigma^2}+\frac{\alpha_2N^{2/3}}{\sigma^2}-\frac{\alpha_3N^{1/3}}{\sigma}+\frac{\alpha_4 N}{\sigma}+\frac{R_1}{2027025}\,\exp\left(-\frac{R_1^2}{2\sigma^2}\right)\Biggl[-\sqrt{\frac{2}{\pi}}\,V_0\,\frac{F_1(R_1,\sigma)}{\sigma^{15}}\nonumber\\
&&+2\sqrt{2\pi}\,n_0\,\frac{K_1(R_1,R_2,\sigma)}{\sigma^{11}}\Biggr]-\frac{R_2}{2027025}\,\exp\left(-\frac{R_2^2}{2\sigma^2}\right)\Biggl[-\sqrt{\frac{2}{\pi}}\,V_0\,\frac{F_2(R_2,\sigma)}{\sigma^{15}}+2\sqrt{2\pi}\,n_0\,\frac{K_2(R_1,R_2,\sigma)}{\sigma^{11}}\Biggr]\nonumber\\
&&+\erf\left(\frac{R_1}{\sqrt{2\sigma}}\right)\Biggl[V_0-\frac{2\pi n_0}{3}K_3(R_1,R_2,\sigma)\Biggr]-\erf\left(\frac{R_2}{\sqrt{2\sigma}}\right)\Biggl[V_0-\frac{2\pi n_0}{3}K_4(R_1,R_2,\sigma)\Biggr].
\end{eqnarray}}

\bibliography{bibfile4}

\end{document}